# Universal classical optical computing inspired by quantum information process

Yifan Sun, Qian Li, Ling-Jun Kong, Jiangwei Shang and Xiangdong Zhang*

*Key Laboratory of advanced optoelectronic quantum architecture and measurements of Ministry of Education, Beijing Key Laboratory of Nanophotonics & Ultrafine Optoelectronic Systems, School of Physics, Beijing Institute of Technology, 100081 Beijing, China.*

**Author to whom any correspondence should be addressed: zhangxd@bit.edu.cn*

**Quantum computing has attracted much attention in recent decades, since it is believed to solve certain problems substantially faster than traditional computing methods. Theoretically, such an advance can be obtained by networks of the quantum operators in universal gate sets, one famous example of which is formed by CNOT gate and single qubit gates. However, realizing a device that performs practical quantum computing is tricky. This is because it requires a scalable qubit system with long coherence time and good controls, which is harsh for most current platforms. Here, we demonstrate that the information process based on a relatively stable system---classical optical system, can be considered as an analogy of universal quantum computing. By encoding the information via the polarization state of classical beams, the optical computing elements that corresponds to the universal gate set are presented and their combination for a general information process are theoretically illustrated. Taking the analogy of two-qubit processor as an example, we experimentally verify that our proposal works well. Considering the potential of optical system for reliable and low-energy-consuming computation, our results open a new way towards advanced information processing with high quality and efficiency.**

The fundamental idea of optical computing, or optical information processing, is based on the benefits of encoding information by light. Using the properties of fast-propagation and coherence, the encoded information is expected to be processed by a device with low heat generation, high degree of parallelism and fast calculation speed[1–6]. During the past 60 years, a series of achievements on optical computing have been done, including joint transform correlator[7], vector matrix multiplier[8,9], energy-conserving gates and circuits[10], etc. However, several important problems are still to be solved. For example, an efficient nonlinear element for optical switching is missing[1,2,11]. It is the key ingredient in the optical digital computing devices which are the mimics of the current electronic computers[3,12.] In recent years, the main research interests of the optical computing have turned to developing the devices for special tasks, and two outstanding works that follow the spirit have been completed. One is the analog optical computing based on metamaterial[4,11,13–15]. In contrast to the digital optical computing scheme, analog optical computing employs the light amplitude distribution as the information carrier, and the engineered photonic structures with certain permittivity and permeability as the modulation approach. Using such a framework, several kinds of application have been realized, such as the calculation of differentiation[16,17], integration[18], Laplace operator[19,20], solutions to linear equations[21], etc. The other refers to optical Ising machines[22–27]. Those machines imitate the networks of interaction magnetic spins by optical setups such as correlated optical parametric oscillators[23–25], bulk optics with spatial light modulators[27], or etc., aiming to solve complex optimization problems. Although the full potential of the Ising machines remains to be explored, their speed-ups over conventional digital computers in several cases have already been demonstrated.

Nevertheless, a blue print for building a universal optical computing device is absent[28]. The investigation on this kind of blue print is important from two aspects. Firstly, assessing the universality of an optical computing system is a theoretical treatment for knowing the upper limit of its computing capability. For example, if an optical system could perform universal computing, it would be powerful enough to calculate all possible computable functions. Secondly, the discussion of the connection of the optical modules for universal computing would also specify the roadmap to large-scale devices applicable for practical issues. One good example of the universal computing model that promotes the relevant researches is quantum computing. Early exploration on universal quantum computing is started by D. Deutsch[29–31], and further explained by D. DiVincenzo[32] and A. Barenco[33] using simple universal gate sets. Based on that foundation, a series of quantum algorithms have been proposed[34–38], including the famous Shor's integer factoring algorithm[36] and Grover's searching algorithm[37]. Also, such foundation settles the direction of the experimental quantum computing research which now concentrates on optimizing the quality of the gate circuits generated by the universal gate sets, and lots of exciting work have been accomplished[39–44].

Inspired by quantum computing strategy, we propose a universal optical computing scheme that can be considered as an analogy of the universal quantum computing, indicating the upper limit of the capability of optical computing to be as powerful as that of quantum computing. Specifically, the operations on light equivalent to the quantum universal gate set are given. Based on our theoretical consideration, we present an experimental verification on the analogy of the two-qubit processor. In the following, we firstly establish the theory of our universal optical computing.

## Universal computing based on the states of classical beams

A physical system for computation requires two basic pillars. One is a mapping of the states of the system to the information that is to be computed. The other is the well-defined control on the system, guaranteeing that the final state after the control would give the results of certain computational tasks. For example, in the theory of quantum computing, a two-level state of a system, or a qubit, is employed to represent a bit of information and the unitary operators on them are the required controls (see supporting material S1 for more details). The system that performs such a computing strategy must contain the degree of freedom for encoding a qubit and can define the unitary controls on it. One of the candidates is the optical system. An optical field possesses a rich degrees of freedom for information encoding, some of which have a close relation with qubits[45–48]. The polarization state of a light field in a two-dimensional plane, for instance, is a well-defined two-level state, no matter whether the light is in the quantum regime or the classical regime. The two-level state based on the polarization of a classical light field can be denoted by

$$|\boldsymbol{E}) = c_0|\boldsymbol{h}) + c_1|\boldsymbol{v}). \tag{1}$$

Here, we use a slightly modified *quantum bra-ket* notation to denote the classical two-level state, which can be considered as the counterpart of a qubit and is termed by a *cebit*[45]. $|\boldsymbol{h})$ and $|\boldsymbol{v})$ denote the basis of the state, referring to the horizontal and vertical polarization state respectively. Complex number $c_0$ and $c_1$ are the projection of cebit state $|\boldsymbol{E})$ on $|\boldsymbol{h})$ and $|\boldsymbol{v})$ respectively, denoted by $c_0 = (\boldsymbol{h}|\boldsymbol{E})$ and $c_1 = (\boldsymbol{v}|\boldsymbol{E})$ with the constrain $|c_0|^2 + |c_1|^2 = 1$. Like the bra-ket notation, the complex conjugate of $|\boldsymbol{E})$ can be denoted by $(\boldsymbol{E}|$. Therefore, the counterpart of density operator can be denoted by $|\boldsymbol{E})(\boldsymbol{E}|$. More generally, for the distinct polarization states of different light fields, an $N$-cebit state $|N\boldsymbol{E})$ can be defined by

$$|N\boldsymbol{E}) = \sum_{j_1,j_2,\ldots,j_N=0}^{1} c_{j_1 j_2 \ldots j_N} |j_1)|j_2)\cdots|j_N), \tag{2}$$

where $|j_1)|j_2)\cdots|j_N)$ represents a bunch of correlated polarization states, each of which is either $|\boldsymbol{h})$ or $|\boldsymbol{v})$, and can be measured independently. The correlation information of the state is characterized by the complex coefficient $c_{j_1 j_2 \ldots j_N}$, the subscripts $j_m$ of which is 0 or 1 for integer $m$ ranging from 1 to $N$. $|j_m) = |\boldsymbol{h})$ when $j_m = 0$, and $|j_m) = |\boldsymbol{v})$ when $j_m = 1$. Comparing equations (1) and (2) with the related qubit states (equation (S1) and (S2) in S1 of Supporting material), one can easily certifies the correspondence between them.

There are various types of classical light fields available for encoding the cebits given by equations (1) and (2). Here, we consider a multi-mode polarized beam for encoding a cebit. The expression of the beam field is given by

$$\boldsymbol{E}(\boldsymbol{r},t) = \sum_{k=1}^{P} f_k(\boldsymbol{r},t)\boldsymbol{p}_k(\boldsymbol{r}), \tag{3}$$

where $\boldsymbol{r}$ and $t$ are the transverse coordinate and time. $\{f_k(\boldsymbol{r},t)\}$ is a set of normalized orthonormal modes in time domain, or spatial domain, or others, under the condition $\int f_{k_1}(\boldsymbol{r},t) \dots f_{k_P}(\boldsymbol{r},t) d\Omega = \delta_{k_1,\dots,k_P}$. $\Omega$ is the parameter of the domain. $\boldsymbol{p}_k(\boldsymbol{r})$ is the polarization vector field of $f_k$. $P$ is the total mode number. Then, $c_0$ and $c_1$ in equation (1) can be given by the projection of equation (3) on the horizontal and vertical polarization state. An example of the setup for obtaining projection is shown in figure 1(a). A local oscillator (LO) $\boldsymbol{E}^{LO}$ is introduced whose polarization state $\boldsymbol{e}$ ($\boldsymbol{h}$, $\boldsymbol{v}$, or their superpositions) is set to be the direction of projection. Then, $\boldsymbol{E}^{LO}$and beam $\boldsymbol{E}$ of a single cebit interferes at the beam splitter (BS in Fig. 1(a)) the outputs are collected by mode-revolved detectors D1 and D2. The real part of $(\boldsymbol{e}|\boldsymbol{E})$ can be obtained by measuring the difference of the signals of D1 and D2 with subtraction device DS. The imaginary part can be measured in the same way, except for adding a $\pi/2$ phase shift to $\boldsymbol{E}^{LO}$.

More generally, an $N$-cebit state can be encoded by $N$ distinguishable beams of the same form with equation (3). Then, $c_{j_1 j_2 \dots j_N}$ of equation (2) can be given by measuring the correlation of the projections of $N$ multi-mode polarized beams, expressed by $c_{j_1 j_2 \dots j_N} \coloneqq [(j_1|(j_2| \cdots (j_N|]|N\boldsymbol{E})$. (For instance, $c_{10\dots 0}$ is given by measuring the correlation of the projection of all the beams on horizontal polarization except for that of the first beam on the vertical, expressed by $[(\boldsymbol{v}|(\boldsymbol{h}| \cdots (\boldsymbol{h}|]|N\boldsymbol{E})$). Such a measurement can be realized by a two-step procedure: 1) measure the local projection of each beam, and 2) multiply the measured signal and integrate the result in the domain of $f_k$. An example of the setup for the correlated measurement is illustrated in figure 1(b). For each beam (denoted by $\cdots \boldsymbol{E}_{n-1}, \boldsymbol{E}_n, \boldsymbol{E}_{n+1} \cdots$), associated LO beams (denoted by $\cdots \boldsymbol{E}_{n-1}^{LO}, \boldsymbol{E}_n^{LO}, \boldsymbol{E}_{n+1}^{LO} \cdots$) are introduced for performing the interferometry setup shown in figure 1(a). After integrating the multiplication of the local projection signals of all the beams, the real (imaginary) part of $c_{j_1 j_2 \dots j_N}$ can be obtained. Such a setup corresponds to the coincidence counters applied to quantum optics experiments. A detailed theoretical analysis of the measurement schemes shown in figure 1(a) and (b) is presented in the part I of Method.

Next, we discuss the operations on the cebits. The universal computing based on cebits requires the well-defined controls on the physical systems such that all cebit states are addressable. Considering the relation between qubits and cebits, those controls are unitary. Therefore, we present the method for implementing arbitrary unitary operations on cebits encoded by multi-mode polarized beams, using the light modulations. We start from the introduction of the setup for two basic operations.

The first one is single cebit operations. Suppose that a multi-mode polarized beam $\boldsymbol{E}_q$ encodes a cebit. The unitary operations on such a single cebit can be denoted by $U_{1E} = R_Y(\xi) R_Z(\eta) R_Y(\zeta)$, where $R_Y(\xi)$ ($R_Z(\eta)$) is the Pauli-$Y$ (Pauli-$Z$) rotation by angle $\xi$ ($\eta$). The operations can be performed by letting $\boldsymbol{E}_q$ pass through a quarter-wave plate (QWP), a half-wave plate (HWP) and a QWP sequentially[49] (also called a Q-H-Q), shown in figure 1(c). The fast axes of them are at angles $\pi/4 + \xi/2$, $-\pi/4 + (\xi + \eta - \zeta)/4$ and $\pi/4 -$ $\zeta/2$ respectively. The output beam is denoted by $\boldsymbol{E}'_q$. Then, the transformation of the cebit encoded by $\boldsymbol{E}_q$ to that encoded by $\boldsymbol{E}'_q$ is $U_{1E}$. Notice that parameter $\xi$, $\eta$ and $\zeta$ are arbitrary real numbers. A theoretical instruction of the relation between the Q-H-Qs and the unitary operations on cebit is given in the part II of Method.

The second one is called a CCX operation. For a two-cebit state $|2\boldsymbol{E}) = c_{00}|\boldsymbol{h})|\boldsymbol{h}) + c_{01}|\boldsymbol{h})|\boldsymbol{v}) + c_{10}|\boldsymbol{v})|\boldsymbol{h}) + c_{11}|\boldsymbol{v})|\boldsymbol{v})$, the CCX operation from the first cebit to the second is defined by $U_{CCX}^{1\to2}|2\boldsymbol{E}) = c_{00}|\boldsymbol{h})|\boldsymbol{h}) + c_{01}|\boldsymbol{h})|\boldsymbol{v}) + c_{11}|\boldsymbol{v})|\boldsymbol{h}) + c_{10}|\boldsymbol{v})|\boldsymbol{v})$, which is an analog of the CNOT gate in quantum circuit model (equation (s4) in supporting material). The superscript $1 \to 2$ means that the first cebit is the control cebit and the second cebit is the target cebit. In the following discussion about a two-cebit process, we hold the convention and omit the superscript. Based on equation (3), the implementation of a two-cebit state requires two beams, denoted by $\boldsymbol{E}_c = \sum_{k_c=1}^{P} f_{k_c}\boldsymbol{p}_{c,k_c}$ and $\boldsymbol{E}_t = \sum_{k_t=1}^{P} f_{k_t}\boldsymbol{p}_{t,k_t}$. The CCX operation can be implemented by inducing the $\boldsymbol{E}_c$-dependent *control* on $\boldsymbol{E}_t$. An example of the setup for a CCX operation is shown in the left part of figure 1(d). For the ease of applying the setup in the following subfigures, we use a blue box with dashed edges to represent the whole setup, indicated by the equals sign in figure 1(d). In such a setup, the beam $\boldsymbol{E}_c$ of the control cebit is split into two parts by a BS. One part is used as the output of the beam of the control cebit. The other is measured by a mode splitter (MS)[50] and a detector $D_M$, which helps to record the coefficients $p_{c,k_c}^H = \boldsymbol{h}\cdot\boldsymbol{p}_{c,k_c}$ and $p_{c,k_c}^V = \boldsymbol{v}\cdot\boldsymbol{p}_{c,k_c}$. Then a programed calculator (PR) gives the matrix $M = F_1^+F_2$, and

$$F_1 = \begin{pmatrix} \boldsymbol{p}_c^H & \boldsymbol{O} \\ \boldsymbol{O} & \boldsymbol{p}_c^H \\ \boldsymbol{p}_c^V & \boldsymbol{O} \\ \boldsymbol{O} & \boldsymbol{p}_c^V \end{pmatrix}, F_2 = \begin{pmatrix} \boldsymbol{p}_c^H & \boldsymbol{O} \\ \boldsymbol{O} & \boldsymbol{p}_c^H \\ \boldsymbol{O} & \boldsymbol{p}_c^V \\ \boldsymbol{p}_c^V & \boldsymbol{O} \end{pmatrix}, \tag{4}$$

where $\boldsymbol{p}_c^H = \left(p_{c,1}^H, p_{c,2}^H, \dots, p_{c,P}^H\right)$ and $\boldsymbol{p}_c^V = \left(p_{c,1}^V, p_{c,2}^V, \dots, p_{c,P}^V\right)$. $\boldsymbol{O}$ is a zero-vector sharing the same size with $\boldsymbol{p}_c^H$. $F_1^+$ is the pseudo inverse matrix[51] of $F_1$. For the beam $\boldsymbol{E}_t$ of the target cebit, an MS is also applied, together with a spatial light modulator (SLM) acting on each mode and a cylinder lens (CL) for bunching the modes, which implements the transformation of $\boldsymbol{E}_t$ to $\boldsymbol{E}'_t$. The parameter of $\boldsymbol{E}'_t$ is given by $\left(\boldsymbol{p}'^H_t, \boldsymbol{p}'^V_t\right)^T = M\cdot(\boldsymbol{p}_t^H, \boldsymbol{p}_t^V)^T$, where $T$ represents the matrix transpose. Using the above definition of cebits, it can be verified that the transformation of the 2-cebit state encoded by $\boldsymbol{E}_c$ and $\boldsymbol{E}_t$ to that encoded by $\boldsymbol{E}_c$ and $\boldsymbol{E}'_t$ is a CCX operation. Obviously, the setup shown in figure 1(d) for the CCX operation is not optimal. However, due to the convenience of manipulating classical light, the setup works in principle. Besides, the setup can also be simplified in special cases. For example, when $P = 2$, consider two beams $\widetilde{\boldsymbol{E}}_c = f_1\boldsymbol{p}_{c,1} + f_2\boldsymbol{p}_{c,2}$ and $\widetilde{\boldsymbol{E}}_t = f_1(\boldsymbol{h}+\boldsymbol{v})/\sqrt{2} + f_2(\boldsymbol{h}-\boldsymbol{v})/\sqrt{2}$, which are sufficient to implement an arbitrary 2-cebit state. The CCX operation from the cebit encoded by $\widetilde{\boldsymbol{E}}_c$ to that encoded by $\widetilde{\boldsymbol{E}}_t$ can be implemented by a much simpler setup, as shown in figure 1(e). In such a setup, the vertical component of mode $f_2$ of $\widetilde{\boldsymbol{E}}_c$ is phase-shifted by a factor of $\pi$. It can be implemented by letting $\widetilde{\boldsymbol{E}}_c$ pass an MS, picking out the vertical component of $f_2$ by a polarization beam splitter (PBS), and then shifting the phase by a phase shifter (PS). Lastly, a PBS and a CL collect the light modes together, giving the output $\widetilde{\boldsymbol{E}}'_c$. The other beam $\widetilde{\boldsymbol{E}}_t$ is not operated. As the output of the setup shown in figure 1(e), $\widetilde{\boldsymbol{E}}'_c$ and $\widetilde{\boldsymbol{E}}_t$ encode the resultant 2-cebit state of that being operated by a CCX. A theoretical background of the setup in figures 1(d) and 1(e) functioning as a CCX operation is given in the part III of Method.

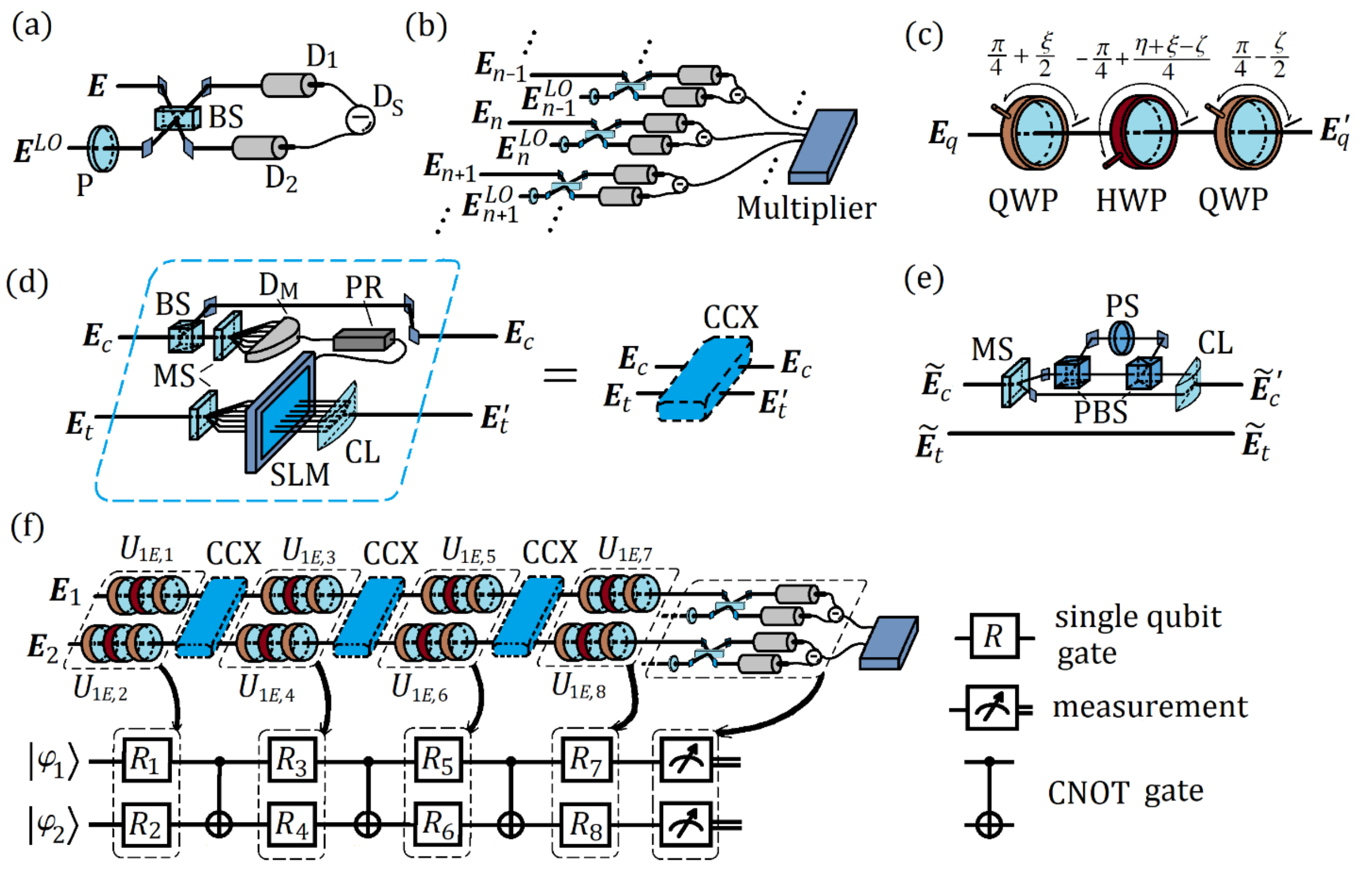

**Fig. 1 | Basic setup of the universal classical optical computing. (a) An example of the local measure, including a BS, a polarizer (P), and LO beam $\boldsymbol{E}^{LO}$. D₁ and D₂ are detectors. Ds is the subtraction device for measuring the difference of the signals recorded by D₁ and D₂. (b) An example of multi-cebit measure. The beams $\cdots \boldsymbol{E}_{n-1}, \boldsymbol{E}_n, \boldsymbol{E}_{n+1} \cdots$ are measured by LO beams $\cdots \boldsymbol{E}^{LO}_{n-1}, \boldsymbol{E}^{LO}_n, \boldsymbol{E}^{LO}_{n+1} \cdots$ using the setup shown by (a). Then, the recorded signals are multiplied by the multiplier. The correlation can be obtained after the integration. (c) Single cebit operation, implemented by the combination of QWP-HWP-QWP (Q-H-Q). The orientation angles of their fast axes are marked in figure. (d) CCX operation, implemented by a BS, an MS, a detector $D_M$, a PR, an SLM, and a CL. The setup is given by the left panel, and a simple denotation of it is given by the right panel. The control cebit is encoded by $\boldsymbol{E}_c$, and the target cebit is encoded by $\boldsymbol{E}_t$. The output of target cebit is encoded by $\boldsymbol{E}'_t$. The output control cebit**

**has no changes. (e) A simplified scheme for CCX considering a special implementation of two-cebits states, including an MS, a PBS, a PS, and a CL. $\widetilde{E}_c$ encodes the control cebit, $\widetilde{E}_t$encodes the target cebit. $\widetilde{E}_c'$ and $\widetilde{E}_t$ are the output. (f) A scheme for an arbitrary 2-cebit operations (upper panel) and its corresponding quantum circuit (lower panel). $U_{1E,1}$~$U_{1E,8}$ are eight single cebit operation implemented by QWPs and HWPs, corresponding to the eight single qubit gates denoted by $R_1$~$R_8$. The relations are marked by arrows.**

Based on the setup of single cebit operation and CCX operation, an arbitrary unitary operation on an $N$-cebit state encoded by $N$ beams can be realized. We start from the two-cebit case. Using three CCXs and eight single cebit operations, an arbitrary unitary operation on a two-cebit state can be expressed by

$$U_{2E} = \left(U_{1E,7} \otimes U_{1E,8}\right)U_{CCX}\left(U_{1E,5} \otimes U_{1E,6}\right)$$
$$\cdot\, U_{CCX}\left(U_{1E,3} \otimes U_{1E,4}\right)U_{CCX}\left(U_{1E,1} \otimes U_{1E,2}\right) \ , \qquad (5)$$

where the $U_{1E,1}$, $U_{1E,2}$,…, and $U_{1E,8}$ denote the arbitrary single cebit operations. Based on the previous discussion, single cebit operations can be implemented by the Q-H-Qs shown in figure 1(c), and the CCX operation can be implemented by the setup shown in figure 1(d) (or figure 1(e) under the constrains). Therefore, the whole setup of equation (5) can be given by connecting the proper Q-H-Qs and the setup for CCXs, as shown in the upper panel of figure 1(f). The two beams output by the setup would encode the resultant two-cebit state of that being operated by a $U_{2E}$. In fact, the setup shown in figure 1(d) (or indicated by equation (5)) is an analogy of the universal 2-qubit processor[52]. We present the circuit of the 2-qubit processor in the lower panel of figure 1(f) and mark the correspondence using arrows. The detailed discussions of the optical setup for equation (5) and the 2-qubit circuit in figure 1(f) are respectively given in S2.1 and S1.1 of Supporting material.

Furthermore, an arbitrary operation on an $N$-cebit state can be constructed. Firstly, we define a special $N$-cebit operation denoted by $M_N R_k$ in a recursive way. When $N = 2$, an $M_2 R_k$ represents a 2-cebit operation defined by $M_2 R_k(\varphi_1, \varphi_2) \coloneqq U_{CCX}(I_2 \otimes U_k(\varphi_2))U_{CCX}(I_2 \otimes U_k(\varphi_1))$, where $I_2$ is a 2-by-2 identity matrix and $U_k(\varphi)$ equals to $R_Y(\varphi)$ $(R_Z(\varphi))$ when $k = y$ $(k = z)$. Such an operation can be implemented by simply two CCX setups and two Q-H-Qs, as shown in figure 2(a). The angle parameters are $\xi = \eta = 0$ and $\zeta = \varphi$ for $U_y(\varphi)$, and $\xi = \zeta = 0$ and $\eta = \varphi$ for $U_z(\varphi)$. Correspondingly, an $M_3 R_k$ represents a 3-cebit operation composed of two CCX operations and two $M_2 R_k$s. Rigorously, an $M_N R_k$ is composed of two CCXs and two $(N-1)$-cebit operation $M_{N-1} R_k$s, expressed by $M_N R_k \colon= U_{CXX}^{1\to N}(I_2 \otimes M_{N-1} R_k)U_{CXX}^{1\to N}(I_2 \otimes M_{N-1} R_k)$. Since one $M_2 R_k$ contains two variables, an $M_3 R_k$ contains four variables, and generally an $M_N R_k$ contains $2^{N-1}$ variables in total. We hide the variables here and below since they are irrelevant to the main conclusion. The optical construction of the recursive relation of $M_N R_k$ is shown in figure 2(b). In such a scheme, the two setups of the CCXs and the two modules for the $M_{N-1}R_k$ are arranged alternatively. The CCX setups are applied for the first beam (control cebit) and the last beam (target cebit). The $M_{N-1}R_k$ modules are applied for all the beams except the first. The ellipsis in the figure indicates the omission of the beams operated by the $M_{N-1}R_k$ and $M_N R_k$ modules. Finally, an arbitrary $M_N R_k$ can be implemented by only Q-H-Qs and CCX setups, if one replaces the $M_{N-1}R_k$ module by the sequence composed of $M_{N-2}R_k$ modules and CCX setups, and then replaces $M_{N-2}R_k$, etc., till $M_2R_k$. In fact, an $M_N R_k$ is an analogy of the quantum multiplexed $R_k$ gate[53]. The circuits of the quantum multiplexed $R_k$ gates are shown in the lower panels of figure 2(a) (the 2-qubit case) and 2(b) (the recursive relation of the $N$-qubit case), and the matrix forms of them are given by equation (s8) and (s9) in S1.2 of Supporting material.

Using $M_N R_k$s, an arbitrary $N$-cebit unitary operation can be expressed. Starting with the 2-cebit operation $U_{2E}$ whose setup is shown in figure 1(f), an arbitrary 3-cebit operation $U_{3E}$ can then be implemented using four $U_{2E}$ setups, two $M_3R_z$ setups, and a $M_3R_y$ setup. Following the same manner, a 4-cebit operation $U_{4E}$ can be implemented using four $U_{3E}$ setups, two $M_4R_z$ setups, and a $M_4R_y$ setup. Strictly, the recursive relation of an arbitrary $N$-cebit unitary operation $U_{NE}$ can be expressed as

$$U_{NE} = \left(I_2 \otimes U_{(N-1)E}\right) \cdot Ro \cdot M_N R_z \cdot Ro^I \cdot \left(I_2 \otimes U_{(N-1)E}\right) \cdot Ro \cdot M_N R_y \cdot Ro^I \cdot \left(I_2 \otimes U_{(N-1)E}\right) \cdot Ro \cdot M_N R_z \cdot Ro^I \cdot \left(I_2 \otimes U_{(N-1)E}\right), \quad (6)$$

where $Ro$ is the operation that swaps the first cebit and rest whole $N-1$ cebits. $Ro^I$ is the inverse of $Ro$. $U_{(N-1)E}$ represents an arbitrary unitary operation on an $(N-1)$-cebit state. Such operations are easy to implement by directly changing the propagation path of the beams, and the optical routers are considered as the approaches for them in our scheme. The setup of equation (6) is shown in the upper panel of figure 2(c). Given $N$ beams that encode an $N$-cebit state, the module of $U_{NE}$ can be given by arranging three $Ro$-$M_N R_k$-$Ro^I$ setups and four $U_{(N-1)E}$ modules alternatively. The $Ro$-$M_N R_k$-$Ro^I$ setup operating on all beams is composed of an $M_N R_k$ setup together with two routers, one of which implements $Ro$ and the other implements $Ro^I$. As shown by the figure, two $Ro$-$M_N R_z$-$Ro^I$ setups and one $Ro$-$M_N R_y$-$Ro^I$ setup are involved. The $U_{(N-1)E}$ modules operate on all the beams except the first, and can be given by a sequence of $U_{(N-2)E}$ modules and the setups of $M_{N-1}R_k$ according to equation (6). The same goes for $U_{(N-2)E}$, etc. In fact, the setup of equation (6) is analogy of the key formula in quantum Shannon decomposition[53]. The circuit of the formula is shown in the lower panel of figure 2(c) (with the correspondence being marked by the arrows), and the matrix form of it is given by equation (s10) and (s11) in S1.2 of Supporting material. Using equation (6) again and again, the module of $U_{NE}$ can be finally implemented by the $U_{2E}$ modules and the $M_2R_k$ setups, which only involves the CCX setups and Q-H-Qs for single cebit operations. Then, a step-by-step manual for building a universal operation on an $N$-cebit state with the two basic setups can be obtained. A theoretical description of the above universal optical computing scheme is given in S2 of Supporting material.

As indicated by our above discussions, the experimental demonstration of the proposal is relatively easy. We take 2-cebit case as an example, and explore firstly the CCX setup and then a universal processor in the following section.

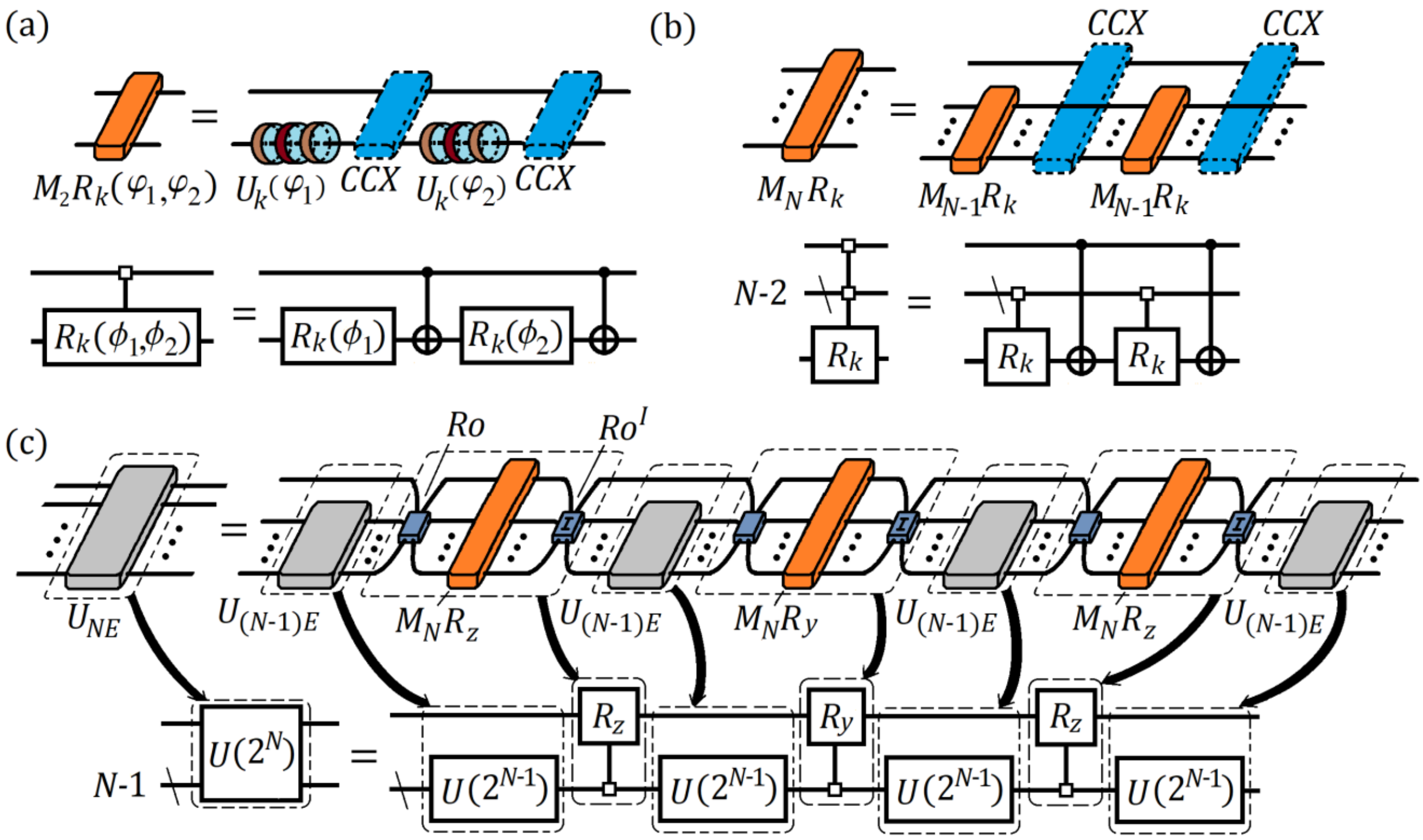

**Fig. 2 | A scheme for building universal cebit operations. (a) The 2-cebit operation $M_2R_k$, realized by two CCXs and two single cebit Pauli rotations. (b) The recursive relation of $M_NR_k$. A $M_NR_k$ can be given by connecting two $M_{N-1}R_k$s and two CCXs as shown by the subfigure. The beams of cebits that go over the CCXs are not affected by the operations. (c) The recursive relation of an arbitrary $N$-cebit operation $U_{NE}$. A $U_{NE}$ can be given by arranging four $U_{(N-1)E}$s and three $M_NR_k$s as shown by the upper panel of the subfigure. The dark blue cuboids (marked by *I*) connecting $U_{(N-1)E}$s and $M_NR_k$s are optical routers which implement the operation $Ro$ ($Ro^I$). As discussed in the main text, the construction corresponds to quantum Shannon decomposition. The key quantum circuits of the decomposition are given below the cebit operations. Especially, we mark the correspondence by arrows in (c). $\phi_1$ and $\phi_2$ are angles of Pauli-rotation $R_k$s in the quantum circuits of (a). The ellipses represent the omission of repetitive notations of beams encoding multi-cebit states.**

**The experimental demonstration of a universal two cebit processor**

We firstly experimentally demonstrate the validity of simplified CCX setup shown in figure 1(e). The experimental scheme is shown in figure 3(a). The scheme is divided into three parts: the input, the CCX operation, and the output. In the input part, the beams of the 2-cebit state are set to contain two spatial modes $f_1$ and $f_2$ respectively. The source of the first beam is a 632.8nm He-Ne laser (ThorlabsHNL210LB), and two BSs (ThorlabsBS016) are employed for producing spatial modes $f_1$ and $f_2$. A polarization sensitive beam displacer (BD) is employed for initially polarizing the laser beam, and the encoding of the cebit states is realized by QWPs and HWPs acting on the two spatial modes. As we mentioned, any 2-cebit state can be generated by the input part (the reason is given by part III of Method). In the CCX part, the operation is realized by an interferometer. It is constructed by two PBSs (ThorlabsPBS201) and a piezoelectric ceramic element M3 for phase modification. This arrangement corresponds to the actions on $\widetilde{\boldsymbol{E}}_c$ shown in figure 1(e). Because we consider the spatial modes which are naturally distinguishable, it is not necessary to apply the MS and the CL.

In the output part, the measurement is performed by realizing the setups shown in figure 1(a) and 1(b). The LO beam is generated by splitting the beam from the previous part. This results in the interferometers shown in the output part of figure 3(a). The two interferometers are used for measuring the two spatial modes, constructed by two BSs and a piezoelectric ceramic element (M1 and M2) individually. The projective direction of the measurement is set by adjusting the polarization state of light in one arm with the wave plates. Although the measurements of two spatial modes are the same so that they can be measured using only one interferometer, we apply two in our experiments so that the visibilities can be improved independently. The light intensity is detected by using a charge coupled devices (CCD, Thorlabs BC106N-VIS/M), and the effective detection area is 8.77 mm □ 6.6 mm. As shown by figure 1(e), the second beam of the cebit is not operated during the procedure. So, we numerically simulated the measurement results instead of actual deeds. To realize it in the experiments, one just needs to generate a beam of two modes from the laser by BSs, adjust the state of polarization by QWPs and HWPs, and measure the beam using at least one interferometer as discussed for the first beam. In our experimental scheme, the results of the correlation are obtained by multiplying the measurement data and simulated data (indicated by the multiplier in figure 3(a)) according to the spatial orthonormal relation, which is equivalent to the integration of the product discussed in the previous section.

Next, we present the results of the CCX experiment. Like the benchmarking of quantum gates, we input a series of 2-cebit states and perform the correlated projection measurement on the states output by the CCX scheme of figure 3(a). The 2-cebit measurement basis is composed of sixteen different tensor products of the states in the set $\{|\boldsymbol{h}\rangle, |\boldsymbol{v}\rangle, (|\boldsymbol{h}\rangle + |\boldsymbol{v}\rangle)/\sqrt{2}, (|\boldsymbol{h}\rangle + i|\boldsymbol{v}\rangle)/\sqrt{2}\}$. By referring to the quantum state tomography theory, the sixteen correlated measurement results can give the density matrix of the output state[54]. For example, by setting the two beams to be $(\boldsymbol{h}+\boldsymbol{v})f_1/\sqrt{2} + (\boldsymbol{h}+\boldsymbol{v})f_2/\sqrt{2}$ and $(\boldsymbol{h}+\boldsymbol{v})f_1/\sqrt{2} + (\boldsymbol{h}-\boldsymbol{v})f_2/\sqrt{2}$, the input 2-cebit state is $(|\boldsymbol{h}\rangle + |\boldsymbol{v}\rangle)|\boldsymbol{h}\rangle/\sqrt{2}$ of the above basis set according to the definition of the cebit measurements. To measure the state output by the CCX part, the two spatial modes of the beam are measured by adjusting the wave plates of single arms of the interferometers according to the above sixteen basis states and multiplying the recorded intensity with the simulated data. Using the method provided by reference[54–56], the density matrix of the output state can be obtained by the sixteen correlations after the normalization, and is shown in figure 1(b). The upper panel is the real part and the lower panel is the imaginary part. Such a density matrix corresponds to the entangled state $(|\boldsymbol{h}\rangle|\boldsymbol{h}\rangle+|\boldsymbol{v}\rangle|\boldsymbol{v}\rangle)/\sqrt{2}$. It indicates the property of a CCX for generating entanglements, corresponding to that of the quantum CNOT gate. Also, we measure the difference between the experimental density matrix and the theoretical one by referring to the quantum state fidelity[54], and find it to be 0.956. Furthermore, we input other fifteen states in the above 2-cebit measurements basis set, and perform the sixteen correlated projection measurements for the outputs respectively, just like what we do when the input is $(|\boldsymbol{h}\rangle+|\boldsymbol{v}\rangle)|\boldsymbol{h}\rangle/\sqrt{2}$. Then, the results of the entire sixteen inputs produce a 16-by-16 data matrix. By referring to the quantum process tomography theory[54,57], one can obtain a 16-by-16 $\chi$ matrix from the data matrix. The entries of the $\chi$ matrix represents the indices of the expansion of a unitary matrix in the basis of Kronecker products of $\{I_2, X, -iY, Z\}$. Because the expansion is unique in the basis, one unitary matrix can be precisely characterized by its $\chi$ matrix. The $\chi$ matrix of our CCX experiment is given in figure S4 of Supporting material which shows an agreement with the theoretical one. The fidelity of implemented CCX operations is found to be 0.989.

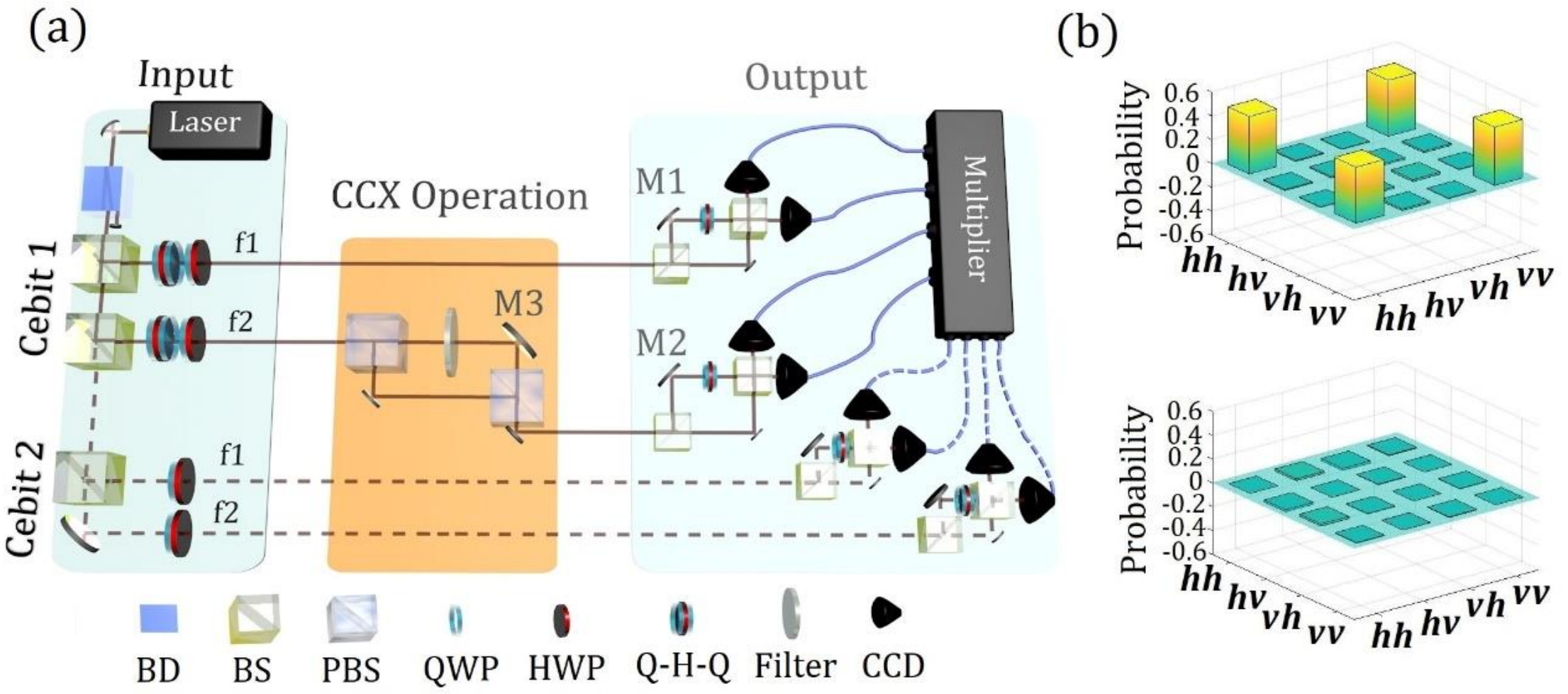

**Fig. 3 | The experiment of CCX operation illustrated by figure 1(e). (a) The optical circuit. The light source is a 632.8 nm He-Ne laser. The beams for encoding the cebit states are generated by BSs and Q-H-Qs. The orange region in (a) marks the CCX operation, composed of two PBSs, a mirror, a piezoelectric ceramic element M3, and a filter for balancing the light intensity. The blue region marks the input and output setup. Piezoelectric ceramic elements M1 and M2 are applied for adjusting the interferometers. The legend is shown in the bottom. (b) The density matrix of the output cebit when the input is $(|\boldsymbol{h}\rangle|\boldsymbol{h}\rangle+|\boldsymbol{v}\rangle|\boldsymbol{h}\rangle)/\sqrt{2}$. The upper (lower) panel shows the real (imaginary) part. The**

**probability here is defined by the normalized the correlation intensity.**

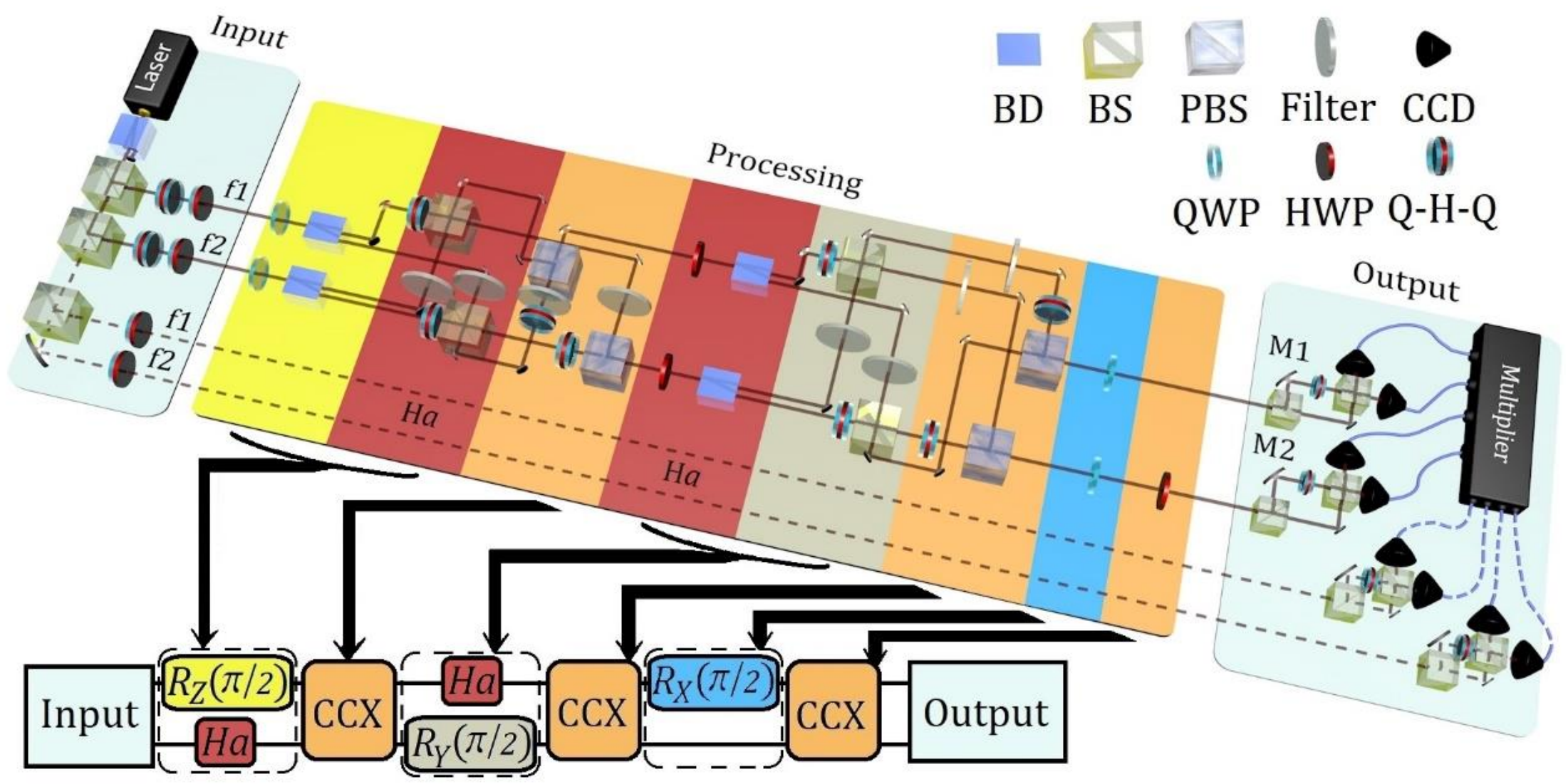

**Fig. 4 | A setup for 2-cebit universal processing. The optical circuit is shown in the middle. The main elements are the same with the CCX experiment, and the filters are also for balancing the light intensity. An instruction of the function of each part is given below, marked by colors and arrows. Especially, the regions for Hadamard operations are noted by $Ha$, for distinguishing them from the other adjacent modules. The piezoelectric ceramic elements M1 and M2 in the output region are also for adjusting the interferometers.**

Based on the setup of the CCX, the experimental scheme for the universal 2-cebit processor is shown in figure 4. The entire setup is also composed of three parts: the input, the processing, and the output. The input and the output are the same with the above CCX experiment. In the processing part, we realize one example of $U_{2E}$ operations defined by equation (5). The eight single cebit operations are set to be $U_{1E,1} = R_Z(\pi/2)$, $U_{1E,2} = Ha$, $U_{1E,3} = Ha$, $U_{1E,4} = R_Y(\pi/2)$, $U_{1E,5} = R_X(\pi/2)$, and $U_{1E,6} = U_{1E,7} = U_{1E,8} = I_2$. $Ha$ is the Hadamard operation for cebits. A general setup for $U_{2E}$ has been demonstrated in figure 1(f). Here, we apply the simplified CCX setup, so the whole implementation of $U_{2E}$ is slightly different from that in figure 1(f). For the operations on the first cebit, they are implemented by Q-H-Qs which is the same as discussed in the previous section. Specifically, the $R_Z(\pi/2)$ and the $Ha$ on the first cebit are implemented by a QWP at 0 rad and an HWP at $\pi/8$ respectively acting on both spatial modes, shown by the yellow and the right dark-red region of figure 4. The $R_X(\pi/2)$ can be implemented based on the equation $R_X(\pi/2) = Ha \cdot R_Z(\pi/2) \cdot Ha$, while here we simplify it to merely modifying the spatial modes by a QWP at $\pi/4$, as shown by the blue region of figure 4. For the operations on the second cebit, they are implemented differently. Due to the requirements of the simplified CCX setup, the second beam must be transformed to $(\boldsymbol{h}+\boldsymbol{v})f_1/\sqrt{2}+$

$(\boldsymbol{h}-\boldsymbol{v})f_2/\sqrt{2}$ before being input to the setup. At the same time, the first beam must also be modified so that the 2-cebit state encoded by them remains the same. Therefore, by keeping the second beam to be $(\boldsymbol{h}+\boldsymbol{v})f_1/\sqrt{2}+(\boldsymbol{h}-\boldsymbol{v})f_2/\sqrt{2}$, the operations on the second cebit can be effectively implemented by changing the physical state of the first beam. Specifically, the Hadamard on the second cebit is effectively performed by modifying the first beam according to $(p_{1,1}^{H}\boldsymbol{h}+p_{1,1}^{V}\boldsymbol{v})f_1+(p_{1,2}^{H}\boldsymbol{h}+p_{1,2}^{V}\boldsymbol{v})f_2 \to [(p_{1,1}^{H}+p_{1,2}^{H})\boldsymbol{h}/\sqrt{2}+(p_{1,1}^{V}+p_{1,2}^{V})\boldsymbol{v}/\sqrt{2}]f_1+[(p_{1,1}^{H}-p_{1,2}^{H})\boldsymbol{h}/\sqrt{2}+(p_{1,1}^{V}-p_{1,2}^{V})\boldsymbol{v}/\sqrt{2}]f_2$. Such a modification can be implemented by an interferometer composed of two BSs and several phase-shift elements, shown by the setup of the left dark-red region in figure 4. Similarly, $R_Y(\pi/2)$ on the second cebit is effectively performed by modifying the first beam according to $(p_{1,1}^{H}\boldsymbol{h}+p_{1,1}^{V}\boldsymbol{v})f_1+(p_{1,2}^{H}\boldsymbol{h}+p_{1,2}^{V}\boldsymbol{v})f_2 \to [(p_{1,1}^{H}+p_{1,2}^{H})\boldsymbol{h}/\sqrt{2}+(p_{1,1}^{V}+p_{1,2}^{V})\boldsymbol{v}/\sqrt{2}]f_1-[(p_{1,1}^{H}-p_{1,2}^{H})\boldsymbol{h}/\sqrt{\mathbf{2}}+(p_{1,1}^{V}-p_{1,2}^{V})\boldsymbol{v}/\sqrt{2}]f_{\mathbf{2}}$. Such a modification can also be implemented by an interferometer like the one for the Hadamard on the second cebit, shown by the setup of the gray region in figure 4. Conclusively, the $U_{2E}$ operation is implemented by modifying the first beam with the above setups following the order illustrated by figure 4. The second beam in such a scheme is also not operated except the measurements. So, like the CCX experiment, we also use numerical results instead. The procedure is indicated by the dashed lines in the setup of figure 4.

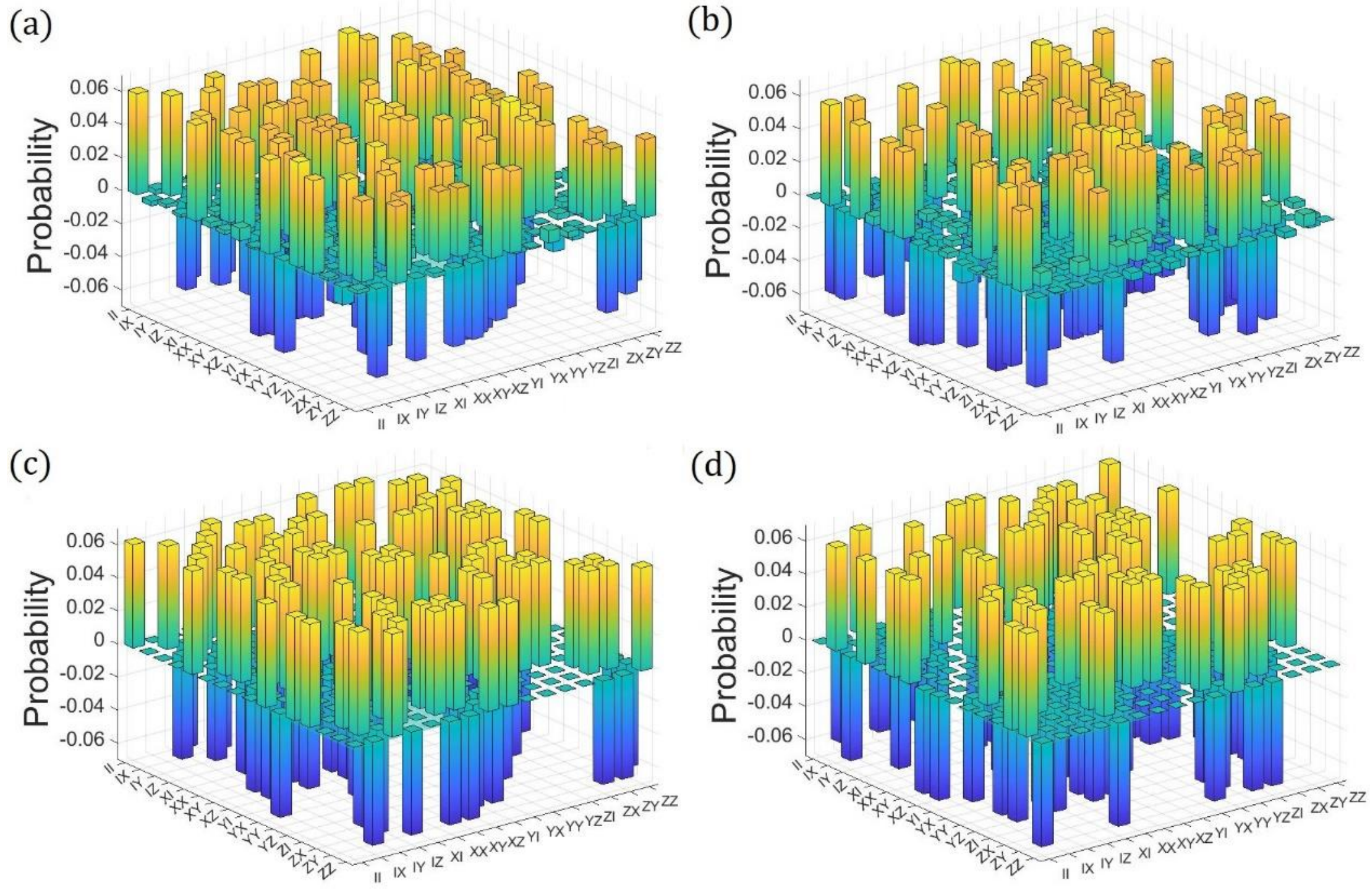

**Fig. 5 | The process tomography results of the setup shown in figure 4. (a) and (b) are the real part and imaginary part of $\chi$ matrix obtained by experimental data. (c) and (d) are the corresponding theoretical results. The probability is also defined by the normalized the correlation intensity.**

The evaluation of the 2-cebit processor is like what we do for CCX experiment. We input the sixteen 2-cebit states, the distinct single cebit state of which is chosen from the set $\{|\boldsymbol{h}\rangle, |\boldsymbol{v}\rangle, (|\boldsymbol{h}\rangle + |\boldsymbol{v}\rangle)/\sqrt{2}, (|\boldsymbol{h}\rangle + i|\boldsymbol{v}\rangle)/\sqrt{2}\}$. For each output, we perform corelated projection measure using the same set with that for CCX experiment. The resultant 16-by-16 data matrix can give the $\chi$ matrix by referring to the quantum process tomography. The result of $\chi$ matrix is graphed by figure 5, characterizing the operation we implement by the scheme shown in figure 4. Figure 5(a) and 5(b) are the real and imaginary part of $\chi$ obtained by experimental data. The theoretical $\chi$ matrix of implemented operation $U_{CCX}(R_X(\pi/2) \otimes I_2)U_{CCX}(Ha \otimes I_2)U_{CCX}(R_Z(\pi/2) \otimes Ha)$ is given by Figure 5(c) and 5(d), which is comparable with experimental results given by 5(a) and 5(b). The fidelity is turned out to be 0.965, sufficient for a reliable computing. Besides the example of $U_{2E}$ we consider here, other 2-cebit computation can also be performed by simply changing the orientations of the fast axes of HWPs and QWPs and the elements of interferometers for the effective operations on the second cebit. A thorough calculation of the whole experimental scheme is given in S3 of Supporting material.

**Discussion and conclusion**

The foundation of our proposal for optical universal computing is two-fold. Firstly, the counterpart of a qubit is defined. It is termed by a cebit, and can be encoded by the corelated classical beams as we proposed in the above. The measurements on cebits also have a good correspondence with the usual setup in quantum experiments for qubits. Secondly, the unitary operations on cebits can be implemented using the conventional method for light modification. Such unitary operations have the same matrix representation with the quantum gates, for instance, the CCX operation and the single cebit operations as mainly discussed by us. The correspondence between the cebit operations and qubit gates also indicated that the cost of the cebit-based computing is comparable with the quantum scheme.

In conclusion, we demonstrate a step-by-step manual for constructing a universal optical computing architecture. It can be considered as an analogy of universal quantum computing. Taking the 2-cebit processor as an example, we experimentally verify our results. We believe that our proposal is important from two aspects. For optical computing, it indicates that the potential of such a reliable and low-energy-consuming strategy for information processing is not fully developed. As we proposed, it can provide an architecture corresponding to the quantum computing which has shown algorithmic advances for several cases. More importantly, our proposal is relatively easy to apply to practical scenarios. The major techniques involved in the scheme are available for current optical processing platforms. Compared with the traditional strategies enabled by the current platforms, our proposal indicates that more advanced strategies would also be realized by them, which are beneficial for achieving the high standard and near-term processors required by the era. Therefore, our results open a new way towards advanced information processing with high quality and efficiency.

## Methods

**I. Theoretical analysis of the measurement scheme for cebits.** We discuss the measurement of the cebit encoded by the multi-mode polarized beam. Using the setup shown by figure 1(a), it is easy to obtain the difference of the intensity recorded by $D_1$ and $D_2$,

$$I_{D1} - I_{D2} = |\boldsymbol{E} + \boldsymbol{E}^{LO}|^2 - |\boldsymbol{E} - \boldsymbol{E}^{LO}|^2 = 2\ \mathrm{Re}\{\boldsymbol{E} \cdot \boldsymbol{E}^{LO*}\}, \tag{7}$$

where Re means the real part, and $*$ represents the complex conjugate. To measure the imaginary part, one just needs to shift the phase of $\boldsymbol{E}^{LO}$ by $\pi/2$. In fact, term $\boldsymbol{E} \cdot \boldsymbol{E}^{LO*}$ has a well correspondence with the projection $\langle e|q\rangle$ of a qubit $|q\rangle$ on $|e\rangle$. For the analogy of an $N$-cebit state, $N$ beams $\boldsymbol{E}_1 \cdots \boldsymbol{E}_{n-1}, \boldsymbol{E}_n, \boldsymbol{E}_{n+1} \cdots \boldsymbol{E}_N$ are introduced. The measurement of the $N$ beams are shown in figure 1(b), expressed by

$$\int (\boldsymbol{e}_1^* \cdot \boldsymbol{E}_1)(\boldsymbol{e}_2^* \cdot \boldsymbol{E}_2) \cdots (\boldsymbol{e}_N^* \cdot \boldsymbol{E}_N)\, d\Omega$$
$$= \int \left(\boldsymbol{e}_1^* \cdot \sum_{k_1=1}^{P} f_{k_1} \boldsymbol{p}_{1,k_1}\right)\left(\boldsymbol{e}_2^* \cdot \sum_{k_2=1}^{P} f_{k_2} \boldsymbol{p}_{2,k_2}\right) \cdots \left(\boldsymbol{e}_N^* \cdot \sum_{k_N=1}^{P} f_{k_N} \boldsymbol{p}_{N,k_N}\right) d\Omega$$
$$= \sum_{k=1}^{P} (\boldsymbol{e}_1^* \cdot \boldsymbol{p}_{1,k})(\boldsymbol{e}_2^* \cdot \boldsymbol{p}_{2,k}) \cdots (\boldsymbol{e}_N^* \cdot \boldsymbol{p}_{N,k})$$
$$= (\boldsymbol{e}_1^* \boldsymbol{e}_2^* \cdots \boldsymbol{e}_N^*) \cdot \sum_{k=1}^{P} (\boldsymbol{p}_{1,k} \boldsymbol{p}_{2,k} \cdots \boldsymbol{p}_{N,k}). \tag{8}$$

The last equality holds due to the property of dyadic tensors $(\boldsymbol{e}_1^* \boldsymbol{e}_2^* \cdots \boldsymbol{e}_N^*)$ and $(\boldsymbol{p}_{1,k} \boldsymbol{p}_{2,k} \cdots \boldsymbol{p}_{N,k})$. Also, Eq. (8) indicates a good correspondence with the projection $\langle e_1|\langle e_2| \cdots \langle e_N|qN\rangle$ of an $N$-qubit state $|qN\rangle$ on $|e_1\rangle|e_2\rangle \cdots |e_N\rangle$. Therefore, the measurement apparatus shown by figure 1(a) and 1(b) can be used to perform the projection of cebits. It also indicates that the cebit space is a Hilbert space.

The conditions for the analogy relation can be strictly given. For the $N$-cebit state defined by Eq. (3), the analogous qubit state can be expressed by,

$$\sum_{j_1, j_2, \ldots, j_N = 0}^{1} q_{j_1 j_2 \ldots j_N} |j_1 j_2 \ldots j_N\rangle, \tag{9}$$

and $q_{j_1 j_2 \ldots j_N}$ is normalized, $q_{j_1 j_2 \ldots j_N} = c_{j_1 j_2 \ldots j_N}$. According to the projection and the measurement given by (8), the $N$ beams that encodes the cebit state corresponding to Eq. (9) must satisfy the equation set

$$\sum_{k=1}^{P}\left(\prod_{d=1}^{N} p_{d,k}^{A_{j_d}}\right) = q_{j_1 j_2 \dots j_N}, \tag{10}$$

where $A_{j_d}$ is set to be $H$ when $j_d = 0$, and to be $V$ when $j_d = 1$. Notice that Eq. (10) is composed of many nonlinear equations. Suppose that we do not require the orthogonality of $f_k$. Similar results can also be obtained, while the expression will be more complicated. Besides, the above way of defining a cebit provide a formulism for simulating a qubit. Other physical systems that can be described by the similar math formulism would also support for an equivalent of a qubit.

**II. The unitary operations of a single cebit based on Q-H-Qs**. The polarization state of a single mode light beam can be arbitrarily modified by QWPs and HWPs[49]. Specifically, a unitary transformation can be parameterized by Euler angles $(\xi, \eta, \zeta)$, expressed by

$$U(\xi,\eta,\zeta) = \exp\left(-i\frac{1}{2}\xi Y\right)\exp\left(-i\frac{1}{2}\eta Z\right)\exp\left(-i\frac{1}{2}\zeta Y\right). \tag{11}$$

The Jones matrices of an HWP and a QWP (the horizontal and vertical polarization states as basis) are given by

$$J_H(\varphi) = \begin{pmatrix} -i\cos 2\varphi & -i\sin 2\varphi \\ -i\sin 2\varphi & i\cos 2\varphi \end{pmatrix}, \text{ and } J_Q(\varphi) = \frac{1}{\sqrt{2}}\begin{pmatrix} 1 - i\cos 2\varphi & -i\sin 2\varphi \\ -i\sin 2\varphi & 1 + i\cos 2\varphi \end{pmatrix}, \tag{12}$$

where $\varphi$ is the orientation angle of the fast axis of the HWP or QWP. The unitary operation on the polarization can be realized by decomposing Eq. (11) into the product of $J_H$ and $J_Q$. One strategy is the Q-H-Q decomposition given by

$$U(\xi,\eta,\zeta) = J_Q\left(\frac{\pi}{4}+\frac{\xi}{2}\right)\cdot J_H\left(-\frac{\pi}{4}+\frac{\xi+\eta-\zeta}{4}\right)\cdot J_Q\left(\frac{\pi}{4}-\frac{\zeta}{2}\right). \tag{13}$$

The multi-mode beams we consider for encoding cebit states can also be operated using a Q-H-Q. Apply Eq. (13) to Eq. (3), one has

$$\begin{aligned} U(\xi,\eta,\zeta)\boldsymbol{E}(\boldsymbol{r},t) &= \sum_{k=1}^{P} f_k(\boldsymbol{r},t)U(\xi,\eta,\zeta)\{p_k^H(\boldsymbol{r})\boldsymbol{h} + p_k^V(\boldsymbol{r})\boldsymbol{v}\} \\ &= \sum_{k=1}^{P} f_k(\boldsymbol{r},t)\{p_k^H(\boldsymbol{r})(u_{00}\boldsymbol{h}+u_{01}\boldsymbol{v}) + p_k^V(\boldsymbol{r})(u_{10}\boldsymbol{h}+u_{11}\boldsymbol{v})\} \\ &= \sum_{k=1}^{P} f_k(\boldsymbol{r},t)\left\{\left(u_{00}p_k^H(\boldsymbol{r}) + u_{10}p_k^V(\boldsymbol{r})\right)\boldsymbol{h} + \left(u_{01}p_k^H(\boldsymbol{r}) + u_{11}p_k^V(\boldsymbol{r})\right)\boldsymbol{v}\right\}, \end{aligned} \tag{14}$$

where $u_{pq}$ is the entry of $U(\xi,\eta,\zeta)$ in the $(p+1)$th row and $(q+1)$th column. $p(q)$ equals to 0 or 1. Assume that the cebit state encoded by $U(\xi,\eta,\zeta)\boldsymbol{E}(\boldsymbol{r},t)$ is $c_0'|h\rangle + c_1'|v\rangle$, then one has

$$c_0' = \sum_{k=1}^{P} f_k(\boldsymbol{r},t)\left(u_{00}p_k^H(\boldsymbol{r}) + u_{10}p_k^V(\boldsymbol{r})\right) = u_{00}c_0 + u_{10}c_1,$$

$$c_1' = \sum_{k=1}^{P} f_k(\boldsymbol{r},t)\left(u_{01}p_k^H(\boldsymbol{r}) + u_{11}p_k^V(\boldsymbol{r})\right) = u_{01}c_0 + u_{11}c_1. \tag{15}$$

Therefore, an arbitrary unitary operation $U_{1E}$ on cebit $c_0|h) + c_1|v)$ encoded by $\boldsymbol{E}(\boldsymbol{r},t)$, can be realized by directly modifying the polarization of the beam with the Q-H-Qs.

**III. The theoretical background of the setup for a CCX operation and the simplification.** We calculate the CCX operation on the beam cebit. Firstly, we consider the 2-cebit case. According to Eq. (10), the parameter equation of $\boldsymbol{E}_c$ and $\boldsymbol{E}_t$ can be given by

$$\begin{pmatrix} \boldsymbol{p}_c^H & \boldsymbol{O} \\ \boldsymbol{O} & \boldsymbol{p}_c^H \\ \boldsymbol{p}_c^V & \boldsymbol{O} \\ \boldsymbol{O} & \boldsymbol{p}_c^V \end{pmatrix} \begin{pmatrix} (\boldsymbol{p}_t^H)^T \\ (\boldsymbol{p}_t^V)^T \end{pmatrix} = \begin{pmatrix} c_{00} \\ c_{01} \\ c_{10} \\ c_{11} \end{pmatrix}. \tag{16}$$

The CCX operation on the 2-cebit state requires that the output beam $\boldsymbol{E}_c'$ equals to $\boldsymbol{E}_c$, and $\boldsymbol{E}_t'$ satisfies the equation

$$\begin{pmatrix} \boldsymbol{p}_c^H & \boldsymbol{O} \\ \boldsymbol{O} & \boldsymbol{p}_c^H \\ \boldsymbol{p}_c^V & \boldsymbol{O} \\ \boldsymbol{O} & \boldsymbol{p}_c^V \end{pmatrix} \begin{pmatrix} (\boldsymbol{p'}_t^H)^T \\ (\boldsymbol{p'}_t^V)^T \end{pmatrix} = \begin{pmatrix} c_{00} \\ c_{01} \\ c_{11} \\ c_{10} \end{pmatrix} = \begin{pmatrix} \boldsymbol{p}_c^H & \boldsymbol{O} \\ \boldsymbol{O} & \boldsymbol{p}_c^H \\ \boldsymbol{O} & \boldsymbol{p}_c^V \\ \boldsymbol{p}_c^V & \boldsymbol{O} \end{pmatrix} \begin{pmatrix} (\boldsymbol{p}_t^H)^T \\ (\boldsymbol{p}_t^V)^T \end{pmatrix}. \tag{17}$$

Using Eq. (4) in the main text, one has

$$F_1 \begin{pmatrix} (\boldsymbol{p'}_t^H)^T \\ (\boldsymbol{p'}_t^V)^T \end{pmatrix} = F_2 \begin{pmatrix} (\boldsymbol{p}_t^H)^T \\ (\boldsymbol{p}_t^V)^T \end{pmatrix}, \text{ or } \begin{pmatrix} (\boldsymbol{p'}_t^H)^T \\ (\boldsymbol{p'}_t^V)^T \end{pmatrix} = F_1^+ F_2 \begin{pmatrix} (\boldsymbol{p}_t^H)^T \\ (\boldsymbol{p}_t^V)^T \end{pmatrix}. \tag{18}$$

Because $F_1$ is not commonly a square matrix, we use the pseudo inverse[51]. Then, as discussed in the main text, such a requirement on output beams can be realized by the setup shown in figure 1(d).

Next, we briefly discuss the CCX operation on two cebits of an $N$-cebit state. Following the part I of Method, suppose that the $n_c$th beam $\boldsymbol{E}_{n_c}$ encodes the control cetbit, the $n_t$th beam $\boldsymbol{E}_{n_t}$ encodes the target cetbit. The requirements on their outputs $\boldsymbol{E}_{n_c}'$ and $\boldsymbol{E}_{n_t}'$ are the same with the above. Therefore, one can obtain the parameters $\boldsymbol{p'}_{n_t}^H$ and $\boldsymbol{p'}_{n_t}^V$ of $\boldsymbol{E}_{n_t}'$ by solving

$$\sum_{k=1}^{P} p_{1,k}^{A_{j_1}} \cdot \ldots \cdot p_{n_c,k}^{H} \cdot \ldots \cdot p'^{H}_{n_t,k} \cdot \ldots \cdot p_{N,k}^{A_{j_N}} = \sum_{k=1}^{P} p_{1,k}^{A_{j_1}} \cdot \ldots \cdot p_{n_c,k}^{H} \cdot \ldots \cdot p_{n_t,k}^{H} \cdot \ldots \cdot p_{N,k}^{A_{j_N}},$$

$$\sum_{k=1}^{P} p_{1,k}^{A_{j_1}} \cdot \ldots \cdot p_{n_c,k}^{V} \cdot \ldots \cdot p'^{H}_{n_t,k} \cdot \ldots \cdot p_{N,k}^{A_{j_N}} = \sum_{k=1}^{P} p_{1,k}^{A_{j_1}} \cdot \ldots \cdot p_{n_c,k}^{V} \cdot \ldots \cdot p_{n_t,k}^{V} \cdot \ldots \cdot p_{N,k}^{A_{j_N}},$$

$$\sum_{k=1}^{P} p_{1,k}^{A_{j_1}} \cdot \ldots \cdot p_{n_c,k}^{H} \cdot \ldots \cdot p'^{V}_{n_t,k} \cdot \ldots \cdot p_{N,k}^{A_{j_N}} = \sum_{k=1}^{P} p_{1,k}^{A_{j_1}} \cdot \ldots \cdot p_{n_c,k}^{H} \cdot \ldots \cdot p_{n_t,k}^{V} \cdot \ldots \cdot p_{N,k}^{A_{j_N}},$$

$$\sum_{k=1}^{P} p_{1,k}^{A_{j_1}} \cdot \ldots \cdot p_{n_c,k}^{V} \cdot \ldots \cdot p'^{V}_{n_t,k} \cdot \ldots \cdot p_{N,k}^{A_{j_N}} = \sum_{k=1}^{P} p_{1,k}^{A_{j_1}} \cdot \ldots \cdot p_{n_c,k}^{V} \cdot \ldots \cdot p_{n_t,k}^{H} \cdot \ldots \cdot p_{N,k}^{A_{j_N}} .$$

(19)

For a sufficiently large $P$, one can find a subspace of the solution under the constrains $\prod_{d=1,d\neq n_c,n_t}^{N} p_{d,k}^{A_{j_d}} = const$. Then, Eqs. (19) can be simplified to

$$\sum_{k=1}^{P} p_{n_c,k}^{H} \cdot p'^{H}_{n_t,k} = \sum_{k=1}^{P} p_{n_c,k}^{H} \cdot p_{n_t,k}^{H} , \sum_{k=1}^{P} p_{n_c,k}^{V} \cdot p'^{H}_{n_t,k} = \sum_{k=1}^{P} p_{n_c,k}^{V} \cdot p_{n_t,k}^{V} ,$$

$$\sum_{k=1}^{P} p_{n_c,k}^{H} \cdot p'^{V}_{n_t,k} = \sum_{k=1}^{P} p_{n_c,k}^{H} \cdot p_{n_t,k}^{V} , \sum_{k=1}^{P} p_{n_c,k}^{V} \cdot p'^{V}_{n_t,k} = \sum_{k=1}^{P} p_{n_c,k}^{V} \cdot p_{n_t,k}^{H} . \quad (20)$$

It is not hard to notice that Eq. (20) is equivalent to Eq. (17). Therefore, the CCX operation on two cebits of an $N$-cebit state can also be implemented using the same setup for the 2-cebit case.

Finally, we calculate the scheme shown by figure 1(e). Using the two beams $\widetilde{\boldsymbol{E}}_c$ and $\widetilde{\boldsymbol{E}}_t$ defined in the main text, a 2-cebit state can be encoded under the condition

$$\frac{1}{\sqrt{2}} \begin{pmatrix} 1 & 1 & 0 & 0 \\ 0 & 0 & 1 & 1 \\ 1 & -1 & 0 & 0 \\ 0 & 0 & 1 & -1 \end{pmatrix} \begin{pmatrix} \tilde{p}_{c,1}^{H} \\ \tilde{p}_{c,2}^{H} \\ \tilde{p}_{c,1}^{V} \\ \tilde{p}_{c,2}^{V} \end{pmatrix} = \begin{pmatrix} c_{00} \\ c_{10} \\ c_{01} \\ c_{11} \end{pmatrix} . \quad (21)$$

It is worth mentioning that Eq. (21) can always be solved. So, $\widetilde{\boldsymbol{E}}_c$ and $\widetilde{\boldsymbol{E}}_t$ can encode an arbitrary 2-cebit state. As discussed above, the CCX operation on the 2-cebit state effectively changes the position of $c_{10}$ and $c_{11}$ in Eq. (21). Then,

$$\begin{pmatrix} c_{00} \\ c_{10} \\ c_{01} \\ c_{11} \end{pmatrix} \xrightarrow{CCX} \begin{pmatrix} c_{00} \\ c_{11} \\ c_{01} \\ c_{10} \end{pmatrix} = \begin{pmatrix} 1 & 0 & 0 & 0 \\ 0 & 0 & 0 & 1 \\ 0 & 0 & 1 & 0 \\ 0 & 1 & 0 & 0 \end{pmatrix} \begin{pmatrix} c_{00} \\ c_{10} \\ c_{01} \\ c_{11} \end{pmatrix} = \frac{1}{\sqrt{2}} \begin{pmatrix} 1 & 0 & 0 & 0 \\ 0 & 0 & 0 & 1 \\ 0 & 0 & 1 & 0 \\ 0 & 1 & 0 & 0 \end{pmatrix} \begin{pmatrix} 1 & 1 & 0 & 0 \\ 0 & 0 & 1 & 1 \\ 1 & -1 & 0 & 0 \\ 0 & 0 & 1 & -1 \end{pmatrix} \begin{pmatrix} \tilde{p}_{c,1}^{H} \\ \tilde{p}_{c,2}^{H} \\ \tilde{p}_{c,1}^{V} \\ \tilde{p}_{c,2}^{V} \end{pmatrix}$$

$$= \frac{1}{\sqrt{2}} \begin{pmatrix} 1 & 1 & 0 & 0 \\ 0 & 0 & 1 & 1 \\ 1 & -1 & 0 & 0 \\ 0 & 0 & 1 & -1 \end{pmatrix} \begin{pmatrix} 1 & 0 & 0 & 0 \\ 0 & 1 & 0 & 0 \\ 0 & 0 & 1 & 0 \\ 0 & 0 & 0 & -1 \end{pmatrix} \begin{pmatrix} \tilde{p}_{c,1}^{H} \\ \tilde{p}_{c,2}^{H} \\ \tilde{p}_{c,1}^{V} \\ \tilde{p}_{c,2}^{V} \end{pmatrix} = \frac{1}{\sqrt{2}} \begin{pmatrix} 1 & 1 & 0 & 0 \\ 0 & 0 & 1 & 1 \\ 1 & -1 & 0 & 0 \\ 0 & 0 & 1 & -1 \end{pmatrix} \begin{pmatrix} \tilde{p}_{c,1}^{H} \\ \tilde{p}_{c,2}^{H} \\ \tilde{p}_{c,1}^{V} \\ -\tilde{p}_{c,2}^{V} \end{pmatrix} .$$

(22)

Therefore, the CCX operation in this case can be implemented by $\tilde{p}_{c,2}^{V} \rightarrow -\tilde{p}_{c,2}^{V}$. The setup is shown by figure 1(e) and the experimental results are presented by figure 3.

## References

1. Ambs, P. Optical Computing: A 60-Year Adventure. *Advances in Optical Technologies* **2010**, 1–15 (2010).
2. Tucker, R. S. The role of optics in computing. *Nature Photon* **4**, 405–405 (2010).
3. Caulfield, H. J. & Dolev, S. Why future supercomputing requires optics. *Nature Photon* **4**, 261–263 (2010).
4. Zangeneh-Nejad, F., Sounas, D. L., Alù, A. & Fleury, R. Analogue computing with metamaterials. *Nat Rev Mater* **6**, 207–225 (2021).
5. Brunner, D., Marandi, A., Bogaerts, W. & Ozcan, A. Photonics for computing and computing for photonics. *Nanophotonics* **9**, 4053–4054 (2020).
6. Lin, X. *et al.* All-optical machine learning using diffractive deep neural networks. *Science* **361**, 1004–1008 (2018).
7. Weaver, C. S. & Goodman, J. W. A Technique for Optically Convolving Two Functions. *Appl. Opt.* **5**, 1248 (1966).
8. Farhat, N. H., Psaltis, D., Prata, A. & Paek, E. Optical implementation of the Hopfield model. *Appl. Opt.* **24**, 1469 (1985).
9. Tamir, D. E., Shaked, N. T., Wilson, P. J. & Dolev, S. High-speed and low-power electro-optical DSP coprocessor. *J. Opt. Soc. Am. A* **26**, A11 (2009).
10. Cohen, E., Dolev, S. & Rosenblit, M. All-optical design for inherently energy-conserving reversible gates and circuits. *Nat Commun* **7**, 11424 (2016).
11. Solli, D. R. & Jalali, B. Analog optical computing. *Nature Photon* **9**, 704–706 (2015).
12. Szöke, A., Daneu, V., Goldhar, J. & Kurnit, N. A. BISTABLE OPTICAL ELEMENT AND ITS APPLICATIONS. *Appl. Phys. Lett.* **15**, 376–379 (1969).
13. Abdollahramezani, S., Hemmatyar, O. & Adibi, A. Meta-optics for spatial optical analog computing. *Nanophotonics* **9**, 4075–4095 (2020).

14. Sihvola, A. Enabling Optical Analog Computing with Metamaterials. *Science* **343**, 144–145 (2014).

15. Kwon, H., Sounas, D., Cordaro, A., Polman, A. & Alù, A. Nonlocal Metasurfaces for Optical Signal Processing. *Phys. Rev. Lett.* **121**, 173004 (2018).

16. Silva, A. *et al.* Performing Mathematical Operations with Metamaterials. *Science* **343**, 160–163 (2014).

17. Zhang, W. & Zhang, X. Backscattering-Immune Computing of Spatial Differentiation by Nonreciprocal Plasmonics. *Phys. Rev. Applied* **11**, 054033 (2019).

18. Zangeneh-Nejad, F. & Khavasi, A. Spatial integration by a dielectric slab and its planar graphene-based counterpart. *Opt. Lett.* **42**, 1954 (2017).

19. Bykov, D. A., Doskolovich, L. L., Bezus, E. A. & Soifer, V. A. Optical computation of the Laplace operator using phase-shifted Bragg grating. *Opt. Express* **22**, 25084 (2014).

20. Guo, C., Xiao, M., Minkov, M., Shi, Y. & Fan, S. Photonic crystal slab Laplace operator for image differentiation. *Optica* **5**, 251 (2018).

21. Barrios, G. A., Retamal, J. C., Solano, E. & Sanz, M. Analog simulator of integro-differential equations with classical memristors. *Sci Rep* **9**, 12928 (2019).

22. Fabre, C. The optical Ising machine. *Nature Photon* **8**, 883–884 (2014).

23. Marandi, A., Wang, Z., Takata, K., Byer, R. L. & Yamamoto, Y. Network of time-multiplexed optical parametric oscillators as a coherent Ising machine. *Nature Photon* **8**, 937–942 (2014).

24. Inagaki, T. *et al.* A coherent Ising machine for 2000-node optimization problems. *Science* **354**, 603–606 (2016).

25. McMahon, P. L. *et al.* A fully programmable 100-spin coherent Ising machine with all-to-all connections. 5.

26. Babaeian, M. *et al.* A single shot coherent Ising machine based on a network of injection-locked multicore fiber lasers. *Nat Commun* **10**, 3516 (2019).

27. Pierangeli, D., Marcucci, G. & Conti, C. Large-Scale Photonic Ising Machine by Spatial Light Modulation. *Phys. Rev. Lett.* **122**, 213902 (2019).

28. We notice that the following work about optical cellular automata has been proposed, but we think that it is hardly applicable for practical tasks. Von Lerber, T., Lassas, M., Le, Q. T. & Küppers, F. Universal All-Optical Computing Based on Interconnected Lasers. in Frontiers in Optics / Laser Science JTu2A.4 (OSA, 2018). doi:10.1364/FIO.2018.JTu2A.4.

29. Quantum theory, the Church–Turing principle and the universal quantum computer. *Proc. R. Soc. Lond. A* **400**, 97–117 (1985).

30. Quantum computational networks. *Proc. R. Soc. Lond. A* **425**, 73–90 (1989).

31. Deutsch, D., Barenco, A. & Ekert, A. Universality in Quantum Computation. *Proc. R. Soc. Lond. A* **449**, 669–677 (1995).

32. DiVincenzo, D. P. Two-bit gates are universal for quantum computation. *Phys. Rev. A* **51**, 1015–1022 (1995).

33. Barenco, A. *et al.* Elementary gates for quantum computation. *Phys. Rev. A* **52**, 3457–3467 (1995).

34. David, D. & Richard, J. Rapid solution of problems by quantum computation. *Proc. R. Soc. Lond. A* **439**, 553–558 (1992).

35. Bernstein, E. & Vazirani, U. Quantum Complexity Theory. *SIAM J. Comput.* **26**, 1411–1473 (1997).

36. Shor, P. W. Polynomial-Time Algorithms for Prime Factorization and Discrete Logarithms on a Quantum Computer. *SIAM J. Comput.* **26**, 1484–1509 (1997).

37. Grover, L. K. A fast quantum mechanical algorithm for database search. in *Proceedings of the twenty-eighth annual ACM symposium on Theory of computing - STOC '96* 212–219 (ACM Press, 1996). doi:10.1145/237814.237866.

38. Moll, N. *et al.* Quantum optimization using variational algorithms on near-term quantum devices. *Quantum Sci. Technol.* **3**, 030503 (2018).

39. Knill, E., Laflamme, R. & Milburn, G. J. A scheme for efficient quantum computation with linear optics. *Nature* **409**, 46–52 (2001).

40. DiCarlo, L. *et al.* Demonstration of two-qubit algorithms with a superconducting quantum processor. *Nature* **460**, 240–244 (2009).

41. Watson, T. F. *et al.* A programmable two-qubit quantum processor in silicon. *Nature* **555**, 633–637 (2018).

42. Qiang, X. *et al.* Large-scale silicon quantum photonics implementing arbitrary two-qubit processing. *Nature Photon* **12**, 534–539 (2018).

43. Wu, Y., Wang, Y., Qin, X., Rong, X. & Du, J. A programmable two-qubit solid-state quantum processor under ambient conditions. *npj Quantum Inf* **5**, 9 (2019).

44. Arute, F. *et al.* Quantum supremacy using a programmable superconducting processor. *Nature* **574**, 505–510 (2019).

45. Spreeuw, R. J. C. Classical wave-optics analogy of quantum-information processing. *Phys. Rev. A* **63**, 062302 (2001).

46. Lee, K. F. & Thomas, J. E. Experimental Simulation of Two-Particle Quantum Entanglement using Classical Fields. *Phys. Rev. Lett.* **88**, 097902 (2002).

47. Sun, Y. *et al.* Non-local classical optical correlation and implementing analogy of quantum teleportation. *Sci Rep* **5**, 9175 (2015).

48. Song, X., Sun, Y., Li, P., Qin, H. & Zhang, X. Bell's measure and implementing quantum Fourier transform with orbital angular momentum of classical light. *Sci Rep* **5**, 14113 (2015).

49. Simon, B. N., Chandrashekar, C. M. & Simon, S. Hamilton's turns as a visual tool kit for designing single-qubit unitary gates. *Phys. Rev. A* **85**, 022323 (2012).

50. Such a device is used for spatially separating the modes. For example, a spectrometer is a mode splitter for frequency. Also, for spatial mode or path mode, a mode splitter is not necessary.

51. *Generalized Inverses*. (Springer-Verlag, 2003). doi:10.1007/b97366.

52. Vatan, F. & Williams, C. Optimal quantum circuits for general two-qubit gates. *Phys. Rev. A* **69**, 032315 (2004).

53.Shende, V. V., Bullock, S. S. & Markov, I. L. Synthesis of quantum-logic circuits. *IEEE Trans. Comput.-Aided Des. Integr. Circuits Syst.* **25**, 1000–1010 (2006).

54.Nielsen, M. A. & Chuang, I. L. *Quantum computation and quantum information*. (Cambridge University Press, 2010).

55.Banaszek, K., D'Ariano, G. M., Paris, M. G. A. & Sacchi, M. F. Maximum-likelihood estimation of the density matrix. *Phys. Rev. A* **61**, 010304 (1999).

56.To obtain the density matrix of a qubit, one just need to measure certain correlations of the quabit state. For cebit state, the only difference is that the correlation measure is replaced by that given in part I of Method.

57.Teo, Y. S., Englert, B.-G., Řeháček, J. & Hradil, Z. Adaptive schemes for incomplete quantum process tomography. *Phys. Rev. A* **84**, 062125 (2011).

# Supporting Information for

# Universal classical optical computing inspired by quantum information process

Yifan Sun, Qian Li, Ling-Jun Kong, Jiangwei Shang and Xiangdong Zhang*

*Key Laboratory of advanced optoelectronic quantum architecture and measurements of Ministry of Education, Beijing Key Laboratory of Nanophotonics & Ultrafine Optoelectronic Systems, School of Physics, Beijing Institute of Technology, 100081 Beijing, China.*

**Author to whom any correspondence should be addressed: zhangxd@bit.edu.cn*

In this supporting material, we present additional details of the main text. In S1, we briefly review the proof of the universal quantum computing. S2 provides supplementary calculation of the universal optical computing scheme. S3 presents a theoretical analysis of our experiments. S4 provides an additional discussion about the generation of the classical beams for encoding cebits.

## S1: Universal quantum computing

In this section, we review the quantum universal computing. The carrier of one bit of information in quantum computing is a theoretical two-level state, i.e., a qubit, denoted by

$$|\psi\rangle = q_0|0\rangle + q_1|1\rangle. \tag{s1}$$

Usually, the computation for a practical task requires multiple qubits, given by

$$|\psi_N\rangle = \sum_{j_1,j_2,\dots,j_N=0}^{1} q_{j_1 j_2 \dots j_N} |j_1 j_2 \dots j_N\rangle. \tag{s2}$$

The universal computing based on qubits is equivalent to operating the qubits to address all the states of the Hilbert space. According to quantum mechanics, the operations satisfy the condition form the unitary group. Therefore, one can perform universal computing if all the unitary operators are implementable. Also, applying the property of Lie group, the implementation of the operators of unitary group can be simplified to the implementation of several basic operators that can generate the whole group. The basic operator sets are often called universal gate sets. Next, we review the proof of one universal gate set composed of CNOT gate and single qubit gates, or more straightforwardly, the blue print for building the universal computing device by two types of gates.

Firstly, we clarify the notations. We use $X$, $Y$, and $Z$ to denote the Pauli matrices. The corresponding rotation are given by

$$R_X(\theta) = \exp\left(-\frac{iX\theta}{2}\right) = \begin{pmatrix} \cos\left(\frac{\theta}{2}\right) & -i\sin\left(\frac{\theta}{2}\right) \\ -i\sin\left(\frac{\theta}{2}\right) & \cos\left(\frac{\theta}{2}\right) \end{pmatrix},$$

$$R_Y(\theta) = \exp\left(-\frac{iY\theta}{2}\right) = \begin{pmatrix} \cos\left(\frac{\theta}{2}\right) & -\sin\left(\frac{\theta}{2}\right) \\ \sin\left(\frac{\theta}{2}\right) & \cos\left(\frac{\theta}{2}\right) \end{pmatrix},$$

$$R_Z(\theta) = \exp\left(-\frac{iZ\theta}{2}\right) = \begin{pmatrix} \exp\left(-\frac{i\theta}{2}\right) & 0 \\ 0 & \exp\left(-\frac{i\theta}{2}\right) \end{pmatrix}. \quad \text{(s3)}$$

For a single qubit, an arbitrary unitary gate can be given by $R_Y(\xi)\, R_Z(\eta)\, R_Y(\zeta)$, where $\xi$, $\eta$, and $\zeta$ are also called Euler angles. The CNOT gate on two qubits can be given by a matrix considering a basis set $\{|00\rangle, |01\rangle, |10\rangle, |11\rangle\}$ (the basis is the same below),

$$U_{CNOT} = \begin{pmatrix} 1 & 0 & 0 & 0 \\ 0 & 1 & 0 & 0 \\ 0 & 0 & 0 & 1 \\ 0 & 0 & 1 & 0 \end{pmatrix}. \quad \text{(s4)}$$

Here, $|ab\rangle = |a\rangle_1 \otimes |b\rangle_2$($a$ and $b$ are 0 or 1), and the first (second) qubit is the control (target) qubit.

## S1.1: The circuit of a universal 2-qubit quantum processor

We discuss the universal 2-qubit quantum processor. Using Cartan decomposition [1], an arbitrary two qubit operator can be given by

$$U_{2q} = (R'_7 \otimes R'_8) \exp[-i(h_1 X \otimes X + h_2 Y \otimes Y + h_3 Z \otimes Z)]\, (R'_1 \otimes R'_2), \quad \text{(s5a)}$$

where

$$R'_1 = R_Y(\gamma_{11}) R_Z(\gamma_{12}) R_Y(\gamma_{13}),\ R'_2 = R_Y(\gamma_{21}) R_Z(\gamma_{22}) R_Y(\gamma_{23}),$$

$$R'_7 = R_Y(\gamma_{31}) R_Z(\gamma_{32}) R_Y(\gamma_{33}), R'_8 = R_Y(\gamma_{41}) R_Z(\gamma_{42}) R_Y(\gamma_{43}), \quad \text{(s5b)}$$

with real numbers $h_1, h_2, h_3$, and $\gamma_{cd}$ ($c = 1,2,7,8;\ d = 1,2,3$). Using CNOT gate and single qubit gates, one can further decompose Eq. (s5) as

$$U_{2q} = (R_7 \otimes R_8) U_{CNOT} (R_5 \otimes R_6) U_{CNOT} (R_3 \otimes R_4) U_{CNOT} (R_1 \otimes R_2), \quad \text{(s6a)}$$

where

$$R_1 = U_{Ha} R'_1, \qquad R_2 = U_{Ha} R_z\left(\frac{\pi}{2}\right) R'_2,$$

$$R_3 = R_z\left(-2h_3 - \frac{\pi}{2}\right) U_{Ha}, \qquad R_4 = R_Y\left(\frac{\pi}{2} + 2h_1\right) U_{Ha},$$

$$R_5 = U_{Ha}, \qquad R_6 = U_{Ha} R_Y\left(-2h_2 - \frac{\pi}{2}\right),$$

$$R_7 = R'_7 R_z\left(-\frac{\pi}{2}\right) U_{Ha}, \qquad R_8 = R'_8 U_{Ha}. \quad \text{(s6b)}$$

$U_{Ha}$ is the Hadamard operator, which can be expressed by $U_{Ha} = XR_Y(\pi/2)$. The circuit is given by figure S1 (corresponding to the lower panel in figure 1(f)).

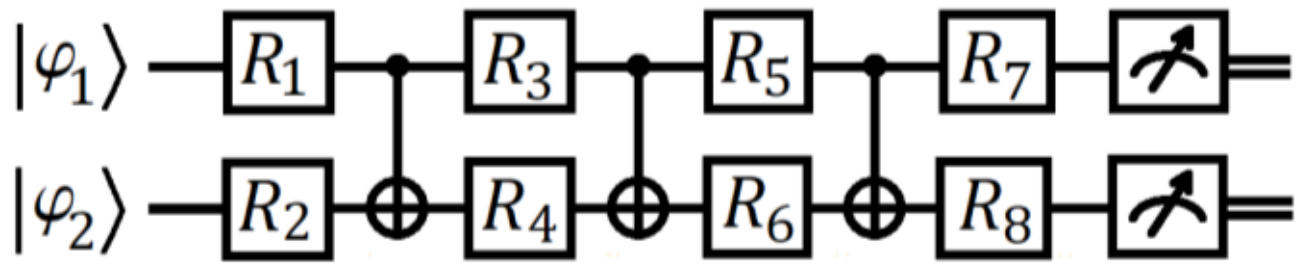

Figure S1: The circuit of a universal two qubit processor.

Eqs. (s6) omits an overall phase factor $e^{i(\pi/4)}$. The matrix form of $U_{2q}$ is given by

$$(R'_7 \otimes R'_8) * \begin{pmatrix} \cos(h_1-h_2)\mathrm{e}^{-ih_3} & 0 & 0 & -i\sin(h_1-h_2)\,\mathrm{e}^{-ih_3} \\ 0 & \cos(h_1+h_2)\,\mathrm{e}^{ih_3} & -i\sin(h_1-h_2)\,\mathrm{e}^{ih_3} & 0 \\ 0 & -i\sin(h_1-h_2)\,\mathrm{e}^{ih_3} & \cos(h_1+h_2)\,\mathrm{e}^{ih_3} & 0 \\ -i\sin(h_1-h_2)\,\mathrm{e}^{-ih_3} & 0 & 0 & \cos(h_1-h_2)\,\mathrm{e}^{-ih_3} \end{pmatrix} * (R'_1 \otimes R'_2), \tag{s7a}$$

where

$$R'_1 \otimes R'_2 = \begin{pmatrix} FC_1 \cdot FC_2 & -FC_1 \cdot FS_2^* & -FS_1^* \cdot FC_2 & FS_1^* \cdot FS_2^* \\ FC_1 \cdot FS_2 & FC_1 \cdot FC_2^* & -FS_1^* \cdot FS_2 & FS_1^* \cdot FC_2^* \\ FS_1 \cdot FC_2 & -FS_1 \cdot FS_2^* & FC_1^* \cdot FC_2 & FC_1^* \cdot FS_2^* \\ FS_1 \cdot FS_2 & FS_1 \cdot FC_2^* & FC_1^* \cdot FS_2 & FC_1^* \cdot FC_2^* \end{pmatrix},$$

$$FS_1 = e^{-i\gamma_{12}} \sin\gamma_{11} \cos\gamma_{13} + \mathrm{e}^{ir} \cos\gamma_{11} \sin\gamma_{13},$$

$$FS_2 = FS_1(\gamma_{11} \to \gamma_{21}, \gamma_{12} \to \gamma_{22}, \gamma_{13} \to \gamma_{23}).$$

$$FC_1 = e^{-i\gamma_{12}} \cos\gamma_{11} \cos\gamma_{13} - \mathrm{e}^{i\gamma_{12}} \sin\gamma_{11} \sin\gamma_{13},$$

$$FC_2 = FC_1(\gamma_{11} \to \gamma_{21}, \gamma_{12} \to \gamma_{22}, \gamma_{13} \to \gamma_{23}). \tag{s7b}$$

$$R'_7 \otimes R'_8 = R'_1 \otimes R'_2(\gamma_{11} \to \gamma_{71}, \gamma_{12} \to \gamma_{72}, \gamma_{13} \to \gamma_{73}, \gamma_{21} \to \gamma_{81}, \gamma_{22} \to \gamma_{82}, \gamma_{23} \to \gamma_{83}). \tag{s7c}$$

In Eq. (s7b), $FS_2$ $(FC_2)$ and $FS_1$ $(FC_2)$ have the same form, except for the replacement $\gamma_{11} \to \gamma_{21}, \gamma_{12} \to \gamma_{22}$, and $\gamma_{13} \to \gamma_{23}$. Eq. (s7c) also means that $R'_7 \otimes R'_8$ and $R'_1 \otimes R'_2$ have the same form, except for the replacement $\gamma_{11} \to \gamma_{71}, \gamma_{12} \to \gamma_{72}, \gamma_{13} \to \gamma_{73}, \gamma_{21} \to \gamma_{81}, \gamma_{22} \to \gamma_{82}$, and $\gamma_{23} \to \gamma_{83}$.

Notice that the circuit given by Eq. (s6) is only an ansatz to the 2-qubit operators. If one chooses $R_1$ to $R_8$ randomly, an arbitrary operator of 2-qubit state can also be obtained using the circuit in figure S1. However, because each single qubit gate is characterized by three parameters, the number of the total parameters of the circuit is 24. The parameter number of a 2-qubit unitary operator is 15. Therefore, randomly choosing $R_1$ to $R_8$ introducing unnecessary parameters for a 2-qubit unitary operator. In fact, the parameter number of Eq. (s6) is 15, which makes it an optimal ansatz.

## S1.2: The circuit of a universal *N*-qubit quantum processor

In this part, we discuss the $N$-qubit case. According to quantum Shannon decomposition [2], an $N$-qubit unitary operator can decomposed into four $(N-1)$-qubit operators and three special $N$-qubit quantum multiplexed $R_k$ gates, denoted by $QM_NR_k$, where $R_k$ refers to Pauli-Y ($k=Y$) or -Z ($k=Z$) rotation. The definition of quantum multiplexed $R_k$ gate can be given from the 2-qubit gate $QM_2R_k$. The matrices form of $QM_2R_k$ can be given by

$$QM_2R_Y(\phi_1,\phi_2)=\begin{pmatrix}\cos\left(\frac{\phi_1+\phi_2}{2}\right) & -\sin\left(\frac{\phi_1+\phi_2}{2}\right) & 0 & 0\\ \sin\left(\frac{\phi_1+\phi_2}{2}\right) & \cos\left(\frac{\phi_1+\phi_2}{2}\right) & 0 & 0\\ 0 & 0 & \cos\left(\frac{\phi_1+\phi_2}{2}\right) & \sin\left(\frac{\phi_1+\phi_2}{2}\right)\\ 0 & 0 & -\sin\left(\frac{\phi_1+\phi_2}{2}\right) & \cos\left(\frac{\phi_1+\phi_2}{2}\right)\end{pmatrix}, \tag{s8a}$$

$$QM_2R_Z(\phi_1,\phi_2)=\begin{pmatrix}e^{-\frac{i}{2}(\phi_1+\phi_2)} & 0 & 0 & 0\\ 0 & e^{\frac{i}{2}(\phi_1+\phi_2)} & 0 & 0\\ 0 & 0 & e^{\frac{i}{2}(\phi_1+\phi_2)} & 0\\ 0 & 0 & 0 & e^{-\frac{i}{2}(\phi_1+\phi_2)}\end{pmatrix}. \tag{s8b}$$

Then, one can obtain $QM_NR_k$ using the following recursive relation

$$\begin{aligned} QM_NR_k &:= \begin{pmatrix} QM_{N-1}R_k & O\\ O & QM_{N-1}R_k^{\dagger}\end{pmatrix}\\ &= \begin{pmatrix} I_{2^{N-1}} & O\\ O & MX_{2^{N-1}}\end{pmatrix}\begin{pmatrix} QM_{N-1}R_k & O\\ O & QM_{N-1}R_k\end{pmatrix}\begin{pmatrix} I_{2^{N-1}} & O\\ O & MX_{2^{N-1}}\end{pmatrix}\begin{pmatrix} QM_{N-1}R_k & O\\ O & QM_{N-1}R_k\end{pmatrix}\\ &= \begin{pmatrix} I_{2^{N-1}} & O\\ O & MX_{2^{N-1}}\end{pmatrix}(I_2\otimes QM_{N-1}R_k)\begin{pmatrix} I_{2^{N-1}} & O\\ O & MX_{2^{N-1}}\end{pmatrix}(I_2\otimes QM_{N-1}R_k), \end{aligned} \tag{s9}$$

where $I_N$ represents the $N$-by-$N$ unity matrix. $O$ is the zero matrix whose size here is $2^{N-1}\times 2^{N-1}$. $MX_{2^{N-1}}$ is a block diagonal matrix with $2^{N-2}$ Pauli-X matrices in the diagonal line. Notice that matrix $diag(I_{2^{N-1}}, MX_{2^{N-1}})$ is a CNOT operator on the first qubit and the last qubit. A $QM_NR_k$ has $2^N$ angles. Here, we only intend to exhibit the logic of a well-known proof. Therefore, the explicit variables of $QM_NR_k$ is irrelevant and we do not reveal them. A circuit representation of Eqs. (s8) and (s9) is given by figure S2.

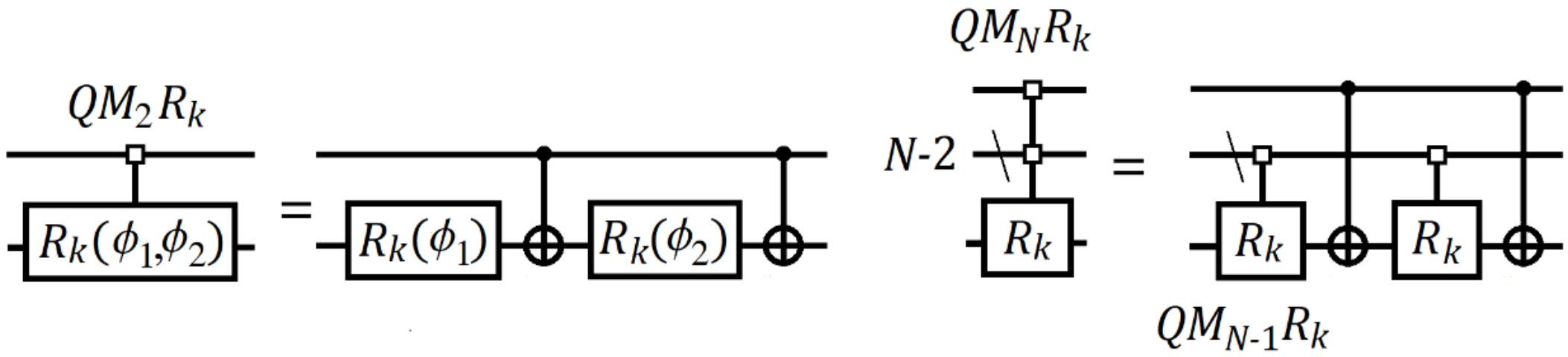

Figure S2: The circuit of $QM_2R_k$ (left) and the recursive relation of $QM_NR_k$ (right).

The decomposition (s9) is based on two properties: 1) the equation of Pauli matrices $X\mathrm{e}^{-i\theta Y}X = \mathrm{e}^{i\theta Y}$ and $X\mathrm{e}^{-i\theta Z}X = \mathrm{e}^{i\theta Z}$, and 2) the product of two $QM_NR_k$ gates is still a $QM_NR_k$. Using $QM_NR_k$ and few-qubit unitary operators, one can obtain the many-qubit unitary operators. The recursive relation can be given in the following way. Considering the cosine-sine decomposition [3,4], a $2^N \times 2^N$ unitary matrix $U_{Nq}$ can be decomposed by

$$U_{Nq} = \begin{pmatrix} A_1 & O \\ O & B_1 \end{pmatrix} \begin{pmatrix} C & -S \\ S & C \end{pmatrix} \begin{pmatrix} A_2 & O \\ O & B_2 \end{pmatrix}, \tag{s10}$$

where $A_1$, $A_2$, $B_1$, and $B_2$ are unitary matrices whose sizes are half of $U_{Nq}$. $C$ and $S$ are diagonal matrices with the same size of $A_1$ and $B_1$, and satisfy $C^2 + S^2 = 1$. The matrix $diag(A_1, B_1)$ (or $diag(A_2, B_2)$) can be decomposed by

$$\begin{pmatrix} A_1 & O \\ O & B_1 \end{pmatrix} = \begin{pmatrix} V & O \\ O & V \end{pmatrix} \begin{pmatrix} D & O \\ O & D^\dagger \end{pmatrix} \begin{pmatrix} W & O \\ O & W \end{pmatrix}, \tag{s11}$$

where $V$ and $W$ are $2^{N-1} \times 2^{N-1}$ unitary matrices. $D$ is a diagonal matrix. $diag(D, D^\dagger)$ represents a $QM_NR_Z$ gate whose operation direction is the opposite of the above definition. Substitute equation (s11) to equation (s10), one can obtain the recursive relation shown by the figure S3.

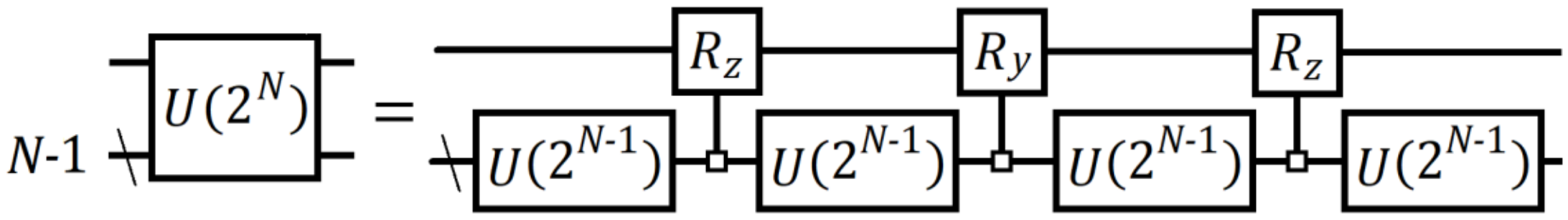

Figure S3: The circuit of the recursive relation of $N$-qubit unitary operators ($U(2^N)$ represent the $2^N$-dimensional unitary group). An $N$-qubit unitary operator can be decomposed into four $(N-1)$-qubit unitary operators and three quantum multiplexed $R_k$ gates.

In conclusion, an arbitrary $N$-qubit unitary operator can be given by recursively applying the circuit shown by figure S3, being generated by 2-qubit gates and 2-qubit multiplexed $R_k$ gates. Therefore, it can be fundamentally constructed by CNOT gates and single qubit gates.

We review the above proof of the universality due to its convenience for illustrating the logic line. Except the quantum Shannon decomposition, there are other proofs. In 1989, D. Deutsch proposed a three-qubit gate and proves it to be universal [5]. Such a gate is termed the Deutsch gate, which is a generalization of Toffoli gate. In 1995, D. DiVincenzo proved that two-qubit gates are universal based on Deutsch gate [6]. At the same year, A. Barenco, et al, propose the universal set composed of CNOT and single qubit gates for the first time, which is also based on Deutsch's work [7]. Besides, according to Solovay-Kiteav theorem, it can be proven that Toffoli gate and Hadamard gate is computationally universal [8]. Finally, Neilson's classic text book shows that arbitrary two-level gates are also universal [9].

# S2: The matrix representation of the information process based on cebit states

In this section, we provide a matrix representation of the key formulas for constructing a universal computing process based on cebit states. We start from the 2-cebit case.

## S2.1: The matrice for 2-cebit universal computing

Based on the cebit defined in the main text, we present matrix form of the cebits,

$$|\boldsymbol{E}) = \begin{pmatrix} c_0 \\ c_1 \end{pmatrix} = \begin{pmatrix} \sum_{k=1}^{P} f_k p_k^H \\ \sum_{k=1}^{P} f_k p_k^V \end{pmatrix}, |N\boldsymbol{E}) = \begin{pmatrix} c_{0\dots00} \\ c_{0\dots01} \\ \vdots \\ c_{1\dots11} \end{pmatrix} = \begin{pmatrix} \sum_{k=1}^{P} p_{1,k}^H \cdots p_{N-1,k}^H p_{N,k}^H \\ \sum_{k=1}^{P} p_{1,k}^H \cdots p_{N-1,k}^H p_{N,k}^V \\ \vdots \\ \sum_{k=1}^{P} p_{1,k}^H \cdots p_{N-1,k}^V p_{N,k}^V \end{pmatrix}. \tag{s12}$$

In part II of Method, we show that the single cebit operations can be implemented by QWPs and HWPs (or Q-H-Qs). The Jones matrices of the QWPs and HWPs are given by Eqs. (12) in the main text. Therefore, the single cebit rotation can be expressed by

$$U_{1E}|\boldsymbol{E}) = J_Q\left(\frac{\pi}{4}+\frac{\xi}{2}\right)J_H\left(\frac{-\pi}{4}+\frac{\xi+\eta-\zeta}{4}\right)J_Q\left(\frac{\pi}{4}-\frac{\zeta}{2}\right)\begin{pmatrix} c_0 \\ c_1 \end{pmatrix}. \tag{s13}$$

Similarly, the CCX operation can also be expressed by

$$U_{CCX}|2\boldsymbol{E}) = \begin{pmatrix} 1 & 0 & 0 & 0 \\ 0 & 1 & 0 & 0 \\ 0 & 0 & 0 & 1 \\ 0 & 0 & 1 & 0 \end{pmatrix} \begin{pmatrix} c_{00} \\ c_{01} \\ c_{10} \\ c_{11} \end{pmatrix} = \begin{pmatrix} c_{00} \\ c_{01} \\ c_{11} \\ c_{10} \end{pmatrix}. \tag{s14}$$

For the 2-cebit states encoded by two classical beams, the universal operations shown in figure 1(f) of the main text can be given by connecting CCX operations with arbitrary Q-H-Qs. Like the quantum case, by properly choosing the QWPs and HWPs, an optimal ansatz corresponding to equation (s6b) can be expressed by

$$U_{1E,1}=J_H\left(\frac{\pi}{8}\right)J_Q\left(\frac{\pi}{4}+\frac{\xi_1}{2}\right)J_H\left(\frac{-\pi}{4}+\frac{\xi_1+\eta_1-\zeta_1}{4}\right)J_Q\left(\frac{\pi}{4}-\frac{\zeta_1}{2}\right),$$

$$U_{1E,2}=J_H\left(\frac{\pi}{8}\right)J_Q\left(\frac{\pi}{4}\right)J_H\left(-\frac{\pi}{8}\right)J_Q\left(\frac{\pi}{4}\right)J_Q\left(\frac{\pi}{4}+\frac{\xi_2}{2}\right)J_H\left(\frac{-\pi}{4}+\frac{\xi_2+\eta_2-\zeta_2}{4}\right)J_Q\left(\frac{\pi}{4}-\frac{\zeta_2}{2}\right),$$

$$U_{1E,3}=J_Q\left(\frac{\pi}{4}\right)J_H\left(-\frac{3\pi}{8}+\frac{c_Z}{2}\right)J_Q\left(\frac{\pi}{4}\right)J_H\left(\frac{\pi}{8}\right),$$

$$U_{1E,4}=J_H\left(-\frac{3\pi}{8}-\frac{c_X}{2}\right)J_Q(-c_X)J_H\left(\frac{\pi}{8}\right),\ U_{1E,5}=J_H\left(\frac{\pi}{8}\right),$$

$$U_{1E,6}=J_H\left(\frac{\pi}{8}\right)J_H\left(-\frac{\pi}{8}+\frac{c_Y}{2}\right)J_Q\left(\frac{\pi}{2}+c_Y\right),$$

$$U_{1E,7}=J_Q\left(\frac{\pi}{4}+\frac{\xi_7}{2}\right)J_H\left(\frac{-\pi}{4}+\frac{\xi_7+\eta_7-\zeta_7}{4}\right)J_Q\left(\frac{\pi}{4}-\frac{\zeta_7}{2}\right)J_Q\left(\frac{\pi}{4}\right)J_H\left(-\frac{3\pi}{8}\right)J_Q\left(\frac{\pi}{4}\right)J_H\left(\frac{\pi}{8}\right),$$

$$U_{1E,8}=J_Q\left(\frac{\pi}{4}+\frac{\xi_8}{2}\right)J_H\left(\frac{-\pi}{4}+\frac{\xi_8+\eta_8-\zeta_8}{4}\right)J_Q\left(\frac{\pi}{4}-\frac{\zeta_8}{2}\right)J_H\left(\frac{\pi}{8}\right). \qquad \text{(s15)}$$

$\xi_m$, $\eta_m$, $\zeta_m$ $(m=1,2,\ 7,\ 8)$, $c_X$, $c_Y$, $c_Z$ are real parameters. Eqs. (s15) are based on the following basic relations,

$$R'_m\rightarrow J_Q\left(\frac{\pi}{4}+\frac{\xi_m}{2}\right)J_H\left(\frac{-\pi}{4}+\frac{\xi_m+\eta_m-\zeta_m}{4}\right)J_Q\left(\frac{\pi}{4}-\frac{\zeta_m}{2}\right),m=1,2,7,8,$$

$$U_{Ha}\rightarrow J_H\left(\frac{\pi}{8}\right),\ R_Y(\theta)\rightarrow J_Q\left(\frac{\pi}{4}\right)J_H\left(-\frac{\pi}{4}-\theta\right)J_Q\left(\frac{\pi}{4}-\frac{\theta}{2}\right),$$

$$R_Z(\theta)\rightarrow J_Q\left(\frac{\pi}{4}\right)J_H\left(-\frac{\pi}{4}+\frac{\alpha}{4}\right)J_Q\left(\frac{\pi}{4}\right). \qquad \text{(s16)}$$

Then, the matrix form of equation (5) in the main text can be given by

$$K_1*\begin{pmatrix}\cos(c_X-c_Y)\mathrm{e}^{-ic_Z} & 0 & 0 & -i\sin(c_X-c_Y)\,\mathrm{e}^{-ic_Z}\\ 0 & \cos(c_X+c_Y)\,\mathrm{e}^{ic_Z} & -i\sin(c_X-c_Y)\,\mathrm{e}^{ic_Z} & 0\\ 0 & -i\sin(c_X-c_Y)\,\mathrm{e}^{ic_Z} & \cos(c_X+c_Y)\,\mathrm{e}^{ic_Z} & 0\\ -i\sin(c_X-c_Y)\,\mathrm{e}^{-ic_Z} & 0 & 0 & \cos(c_X-c_Y)\,\mathrm{e}^{-ic_Z}\end{pmatrix}*K_0,$$

(s17a)

where $K_1$ and $K_0$ can be given by using the forms of equations (s7b) and (s7c) and replacing the variables,

$$K_0=R'_1\otimes R'_2(\gamma_{11}\rightarrow\xi_1,\gamma_{12}\rightarrow\eta_1,\gamma_{13}\rightarrow\zeta_1,\gamma_{21}\rightarrow\xi_2,\gamma_{22}\rightarrow\eta_2,\gamma_{23}\rightarrow\zeta_2),$$

$$K_1=R'_7\otimes R'_8(\gamma_{71}\rightarrow\xi_7,\gamma_{72}\rightarrow\eta_7,\gamma_{73}\rightarrow\zeta_7,\gamma_{81}\rightarrow\xi_8,\gamma_{82}\rightarrow\eta_8,\gamma_{83}\rightarrow\zeta_8). \qquad \text{(s17b)}$$

It can be concluded that the matrix (s17a) has the same form with (s7a), except a replacement of the variables given by $h_1\rightarrow c_X$, $h_2\rightarrow c_Y$, $h_3\rightarrow c_Z$, and Eq. (s17b).

## S2.2: The matrices for decomposing the *N*-cebit unitary operations

Following the discussion in the main text, we firstly give the $M_2R_k(\varphi_1,\varphi_2)$ operation, shown in figure 2(a). Using Eqs. (s16), one has

$$M_2R_Y(\varphi_1,\varphi_2) = U_{CCX}\left\{I_2\otimes\left[J_H\left(-\frac{\pi}{4}-\varphi_1\right)J_Q\left(\frac{\pi}{4}-\frac{\varphi_1}{2}\right)\right]\right\}U_{CCX}\left\{I_2\otimes\left[J_H\left(-\frac{\pi}{4}-\varphi_2\right)J_Q\left(\frac{\pi}{4}-\frac{\varphi_2}{2}\right)\right]\right\}$$

$$= \begin{pmatrix} \cos\left(\frac{\varphi_1+\varphi_2}{2}\right) & -\sin\left(\frac{\varphi_1+\varphi_2}{2}\right) & 0 & 0 \\ \sin\left(\frac{\varphi_1+\varphi_2}{2}\right) & \cos\left(\frac{\varphi_1+\varphi_2}{2}\right) & 0 & 0 \\ 0 & 0 & \cos\left(\frac{\varphi_1+\varphi_2}{2}\right) & \sin\left(\frac{\varphi_1+\varphi_2}{2}\right) \\ 0 & 0 & -\sin\left(\frac{\varphi_1+\varphi_2}{2}\right) & \cos\left(\frac{\varphi_1+\varphi_2}{2}\right) \end{pmatrix},$$

$$M_2R_Z(\varphi_1,\varphi_2) = U_{CCX}\left\{I_2\otimes\left[J_Q\left(\frac{\pi}{4}\right)J_H\left(-\frac{\pi}{4}+\frac{\varphi_1}{4}\right)J_Q\left(\frac{\pi}{4}\right)\right]\right\}U_{CCX}\left\{I_2\otimes\left[J_Q\left(\frac{\pi}{4}\right)J_H\left(-\frac{\pi}{4}+\frac{\varphi_2}{4}\right)J_Q\left(\frac{\pi}{4}\right)\right]\right\} =$$

$$\begin{pmatrix} e^{-\frac{i}{2}(\varphi_1+\varphi_2)} & 0 & 0 & 0 \\ 0 & e^{\frac{i}{2}(\varphi_1+\varphi_2)} & 0 & 0 \\ 0 & 0 & e^{\frac{i}{2}(\varphi_1+\varphi_2)} & 0 \\ 0 & 0 & 0 & e^{-\frac{i}{2}(\varphi_1+\varphi_2)} \end{pmatrix}. \tag{s18}$$

Obviously, equations (s18) are equivalent to equations (s8). Also, the recursive relation of $M_NR_k$ can be given by

$$M_NR_k = U_{CCX}^{1\to N}\begin{pmatrix} M_{N-1}R_k & O \\ O & M_{N-1}R_k \end{pmatrix}U_{CCX}^{1\to N}\begin{pmatrix} M_{N-1}R_k & O \\ O & M_{N-1}R_k \end{pmatrix} = \begin{pmatrix} M_{N-1}R_k & O \\ O & M_{N-1}R_k^{\dagger} \end{pmatrix}, \tag{s20}$$

which corresponds to Eq. (s9). According to Eq. (s20), an $M_NR_k$ also has $2^N$ angle variables, as mentioned in the main text. Due to the exact correspondence between the $M_NR_k$ and the $QM_NR_k$, we can completely follow the mathematical form of the quantum proof so that the specific variables are not important. Therefore, like the previous discussion, we also omit the explicit parameters of the $M_NR_k$. Finally, an $M_NR_k$ for $N$-cebit states can be built by starting from $M_2R_k$ and repetitively using (s20). Then, using cosine-sine decomposition, we can obtain the results given by Eq. (6) in the main text. Such a process has a one-to-one correspondence with that shown in S1.2.

## S3: The theoretical analysis of the experiments

We firstly present the theoretical analysis of the CCX experiments shown by figure 3. The two spatial light modes are generated by the laser and the two BSs. Each mode can be arbitrarily modified by the QWPs and HWPs. Therefore, the expressions of the two spatial modes can be given by $(p_{c,1}^H\boldsymbol{h}+p_{c,1}^V\boldsymbol{v})f_1$ and $(p_{c,2}^H\boldsymbol{h}+p_{c,2}^V\boldsymbol{v})f_2$. Next, the transformations of the two spatial modes are

$$(p_{c,1}^H\boldsymbol{h}+p_{c,1}^V\boldsymbol{v})f_1 \to (p_{c,1}^H\boldsymbol{h}+p_{c,1}^V\boldsymbol{v})f_1,$$

$$(p_{c,2}^H\boldsymbol{h}+p_{c,2}^V\boldsymbol{v})f_2 \xrightarrow{PBS} \left\{\begin{matrix} p_{c,2}^H\boldsymbol{h}f_2 \to p_{c,2}^H\boldsymbol{h}f_2 \\ p_{c,2}^V\boldsymbol{v}f_2 \xrightarrow{M3} -p_{c,2}^V\boldsymbol{v}f_2 \end{matrix}\right\} \xrightarrow{PBS} (p_{c,2}^H\boldsymbol{h}-p_{c,2}^V\boldsymbol{v})f_2. \tag{s21}$$

The curly brackets are used to represent the output light fields propagating along different paths (the same below). As mentioned in the main text, M3 is used for shifting the phase and the filter is used for balancing the intensity. Then, the measurement setup of the first mode provides the projection of the output of (s21)

$$(p_{c,1}^{H}\boldsymbol{h}+p_{c,1}^{V}\boldsymbol{v})f_1 \xrightarrow{BS} \begin{Bmatrix} \frac{1}{\sqrt{2}}(p_{c,1}^{H}\boldsymbol{h}+p_{c,1}^{V}\boldsymbol{v})f_1 \rightarrow \frac{1}{\sqrt{2}}(p_{c,1}^{H}\boldsymbol{h}+p_{c,1}^{V}\boldsymbol{v})f_1 \\ \frac{1}{\sqrt{2}}(p_{c,1}^{H}\boldsymbol{h}+p_{c,1}^{V}\boldsymbol{v})f_1 \xrightarrow{Q-H-Q} \frac{1}{\sqrt{2}}\boldsymbol{e}_c f_1 \end{Bmatrix}$$

$$\xrightarrow{BS} \begin{Bmatrix} \frac{1}{\sqrt{2}}(p_{c,1}^{H}\boldsymbol{h}+p_{c,1}^{V}\boldsymbol{v}+\boldsymbol{e}_c)f_1 \\ \frac{1}{\sqrt{2}}(p_{c,1}^{H}\boldsymbol{h}+p_{c,1}^{V}\boldsymbol{v}-\boldsymbol{e}_c)f_1 \end{Bmatrix} \xrightarrow{CCD\ \&\ \Delta} \frac{1}{\sqrt{2}}\ \mathrm{Re}\left\{\left(p_{c,1}^{H}(\boldsymbol{e}_c^{*}\cdot\boldsymbol{h})+p_{c,1}^{V}(\boldsymbol{e}_c^{*}\cdot\boldsymbol{v})\right)f_1\right\}, \quad \text{(s22)}$$

where $\boldsymbol{e}_c$ represents the projection direction. The measurement setup of the second mode is the same with the first, giving that $\frac{1}{\sqrt{2}}\ \mathrm{Re}\left\{\left(p_{c,2}^{H}(\boldsymbol{e}_c^{*}\cdot\boldsymbol{h})-p_{c,2}^{V}(\boldsymbol{e}_c^{*}\cdot\boldsymbol{v})\right)f_2\right\}$. The imaginary part can be obtained by changing $\boldsymbol{e}_c$ to $i\boldsymbol{e}_c$. The virtual beam is set to be $f_1(\boldsymbol{h}+\boldsymbol{v})/\sqrt{2}$ and $f_2(\boldsymbol{h}-\boldsymbol{v})/\sqrt{2}$. Their projection on $\boldsymbol{e}_t$ is given by $f_1(\boldsymbol{e}_t^{*}\cdot\boldsymbol{h}+\boldsymbol{e}_t^{*}\cdot\boldsymbol{v})/\sqrt{2}$ and $f_2(\boldsymbol{e}_t^{*}\cdot\boldsymbol{h}-\boldsymbol{e}_t^{*}\cdot\boldsymbol{v})/\sqrt{2}$. By numerically multiply the results, one obtains the real part proportional to

$$\begin{aligned}&\left[p_{c,1}^{H}(\boldsymbol{e}_c^{*}\cdot\boldsymbol{h})+p_{c,1}^{V}(\boldsymbol{e}_c^{*}\cdot\boldsymbol{v})\right](\boldsymbol{e}_t^{*}\cdot\boldsymbol{h}+\boldsymbol{e}_t^{*}\cdot\boldsymbol{v})+\left[p_{c,2}^{H}(\boldsymbol{e}_c^{*}\cdot\boldsymbol{h})-p_{c,2}^{V}(\boldsymbol{e}_c^{*}\cdot\boldsymbol{v})\right](\boldsymbol{e}_t^{*}\cdot\boldsymbol{h}-\boldsymbol{e}_t^{*}\cdot\boldsymbol{v})\\ &=(\boldsymbol{e}_c^{*}\boldsymbol{e}_t^{*})\cdot\left[(p_{c,1}^{H}+p_{c,2}^{H})(\boldsymbol{hh})+(p_{c,1}^{H}-p_{c,2}^{H})(\boldsymbol{hv})+(p_{c,1}^{V}-p_{c,2}^{V})(\boldsymbol{vh})+(p_{c,1}^{V}+p_{c,2}^{V})(\boldsymbol{vv})\right]. \quad \text{(s23)}\end{aligned}$$

Here, we also use the dyadic tensor notation. Consider the projection of the input state on $\boldsymbol{e}_c^{*}\boldsymbol{e}_t^{*}$, which is proportional to

$$\begin{aligned}&\left[p_{c,1}^{H}(\boldsymbol{e}_c^{*}\cdot\boldsymbol{h})+p_{c,1}^{V}(\boldsymbol{e}_c^{*}\cdot\boldsymbol{v})\right](\boldsymbol{e}_t^{*}\cdot\boldsymbol{h}+\boldsymbol{e}_t^{*}\cdot\boldsymbol{v})+\left[p_{c,2}^{H}(\boldsymbol{e}_c^{*}\cdot\boldsymbol{h})+p_{c,2}^{V}(\boldsymbol{e}_c^{*}\cdot\boldsymbol{v})\right](\boldsymbol{e}_t^{*}\cdot\boldsymbol{h}-\boldsymbol{e}_t^{*}\cdot\boldsymbol{v})\\ &=(\boldsymbol{e}_c^{*}\boldsymbol{e}_t^{*})\cdot\left[(p_{c,1}^{H}+p_{c,2}^{H})(\boldsymbol{hh})+(p_{c,1}^{H}-p_{c,2}^{H})(\boldsymbol{hv})+(p_{c,1}^{V}+p_{c,2}^{V})(\boldsymbol{vh})+(p_{c,1}^{V}-p_{c,2}^{V})(\boldsymbol{vv})\right]. \quad \text{(s24)}\end{aligned}$$

Following Eq. (s12), one has

$$\begin{pmatrix} p_{c,1}^{H}+p_{c,2}^{H} \\ p_{c,1}^{H}-p_{c,2}^{H} \\ p_{c,1}^{V}-p_{c,2}^{V} \\ p_{c,1}^{V}+p_{c,2}^{V} \end{pmatrix} = U_{CCX} \begin{pmatrix} p_{c,1}^{H}+p_{c,2}^{H} \\ p_{c,1}^{H}-p_{c,2}^{H} \\ p_{c,1}^{V}+p_{c,2}^{V} \\ p_{c,1}^{V}-p_{c,2}^{V} \end{pmatrix}, \quad \text{(s25)}$$

which corresponds to Eq. (22) in the part III of Method. The results of Eq. (s23) correspond to the corelated projection of a quantum state. Therefore, one can obtain the density matrix of the output state using Eq. (s23) and the quantum state tomography theory. The corresponding results are given by figure 3(b). Also, we use the states $|\boldsymbol{h}\rangle|\boldsymbol{h}\rangle$, $|\boldsymbol{h}\rangle|\boldsymbol{v}\rangle$, $|\boldsymbol{h}\rangle|+\rangle$, $|\boldsymbol{h}\rangle|R\rangle$, $|\boldsymbol{v}\rangle|\boldsymbol{h}\rangle$, $|\boldsymbol{v}\rangle|\boldsymbol{v}\rangle$, $|\boldsymbol{v}\rangle|+\rangle$, $|\boldsymbol{v}\rangle|R\rangle$, $|+\rangle|\boldsymbol{h}\rangle$, $|+\rangle|\boldsymbol{v}\rangle$, $|+\rangle|+\rangle$, $|+\rangle|R\rangle$, $|R\rangle|\boldsymbol{h}\rangle$, $|R\rangle|\boldsymbol{v}\rangle$, $|R\rangle|+\rangle$, and $|R\rangle|R\rangle$ as the input, and obtain the process tomography results of the CCX operation shown in figure S4.

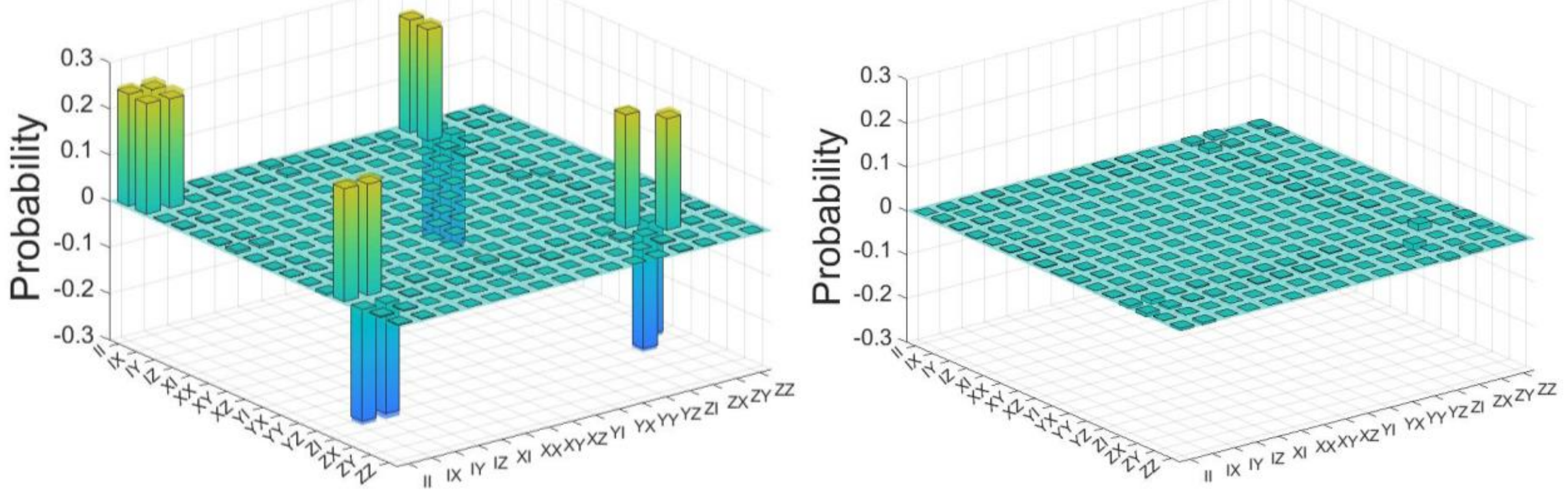

Figure S4: The process tomography result of CCX, represented by the $\chi$ matrix. The left panel gives the real part and the right panel gives the imaginary part. The black frames of the cuboids give the experimental results and the colored bars give the theoretical results.

Next, we present the analysis of the optical setup shown by figure 4 step by step. Like the CCX operation, the two spatial modes generated by the laser and the BSs are $(p_{c,1}^{H}\boldsymbol{h}+p_{c,1}^{V}\boldsymbol{v})f_1$ and $(p_{c,2}^{H}\boldsymbol{h}+p_{c,2}^{V}\boldsymbol{v})f_2$. The virtual beams are also $f_1(\boldsymbol{h}+\boldsymbol{v})/\sqrt{2}$ and $f_2(\boldsymbol{h}-\boldsymbol{v})/\sqrt{2}$. The operations on the beam are shown as follows. (In the above, we use $U_{Ha}$ to denote the quantum Hadamard operator. For distinguishment, we use $Ha$ to denote the Hadamard operation on a cebit blow and in the main text. The matrix form of $U_{Ha}$ and $Ha$ could be the same.)

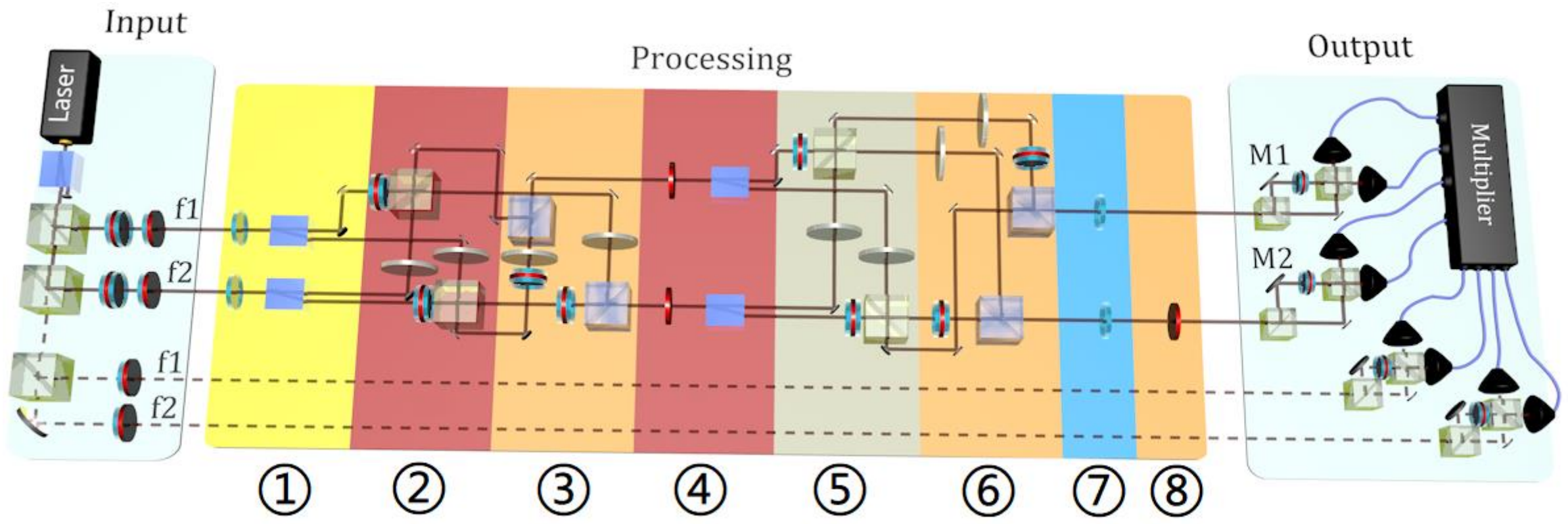

Figure S5: The optical setup shown by figure 4 with number marks.

① $R_z(\pi/2)$ on the first cebit.

Optical setup:

$$(p_{c,1}^{H}\boldsymbol{h}+p_{c,1}^{V}\boldsymbol{v})f_1 \xrightarrow{QWP@0^\circ} \left(\frac{1-i}{\sqrt{2}}p_{c,1}^{H}\boldsymbol{h}+\frac{1+i}{\sqrt{2}}p_{c,1}^{V}\boldsymbol{v}\right)f_1,$$

$$(p_{c,2}^{H}\boldsymbol{h}+p_{c,2}^{V}\boldsymbol{v})f_2 \xrightarrow{QWP@0^\circ} \left(\frac{1-i}{\sqrt{2}}p_{c,2}^{H}\boldsymbol{h}+\frac{1+i}{\sqrt{2}}p_{c,2}^{V}\boldsymbol{v}\right)f_2,$$

Output state:

$$\left[\frac{1-i}{\sqrt{2}}p_{c,1}^{H}\boldsymbol{h}+\frac{1+i}{\sqrt{2}}p_{c,1}^{V}\boldsymbol{v}\right](\boldsymbol{h}+\boldsymbol{v})+\left[\frac{1-i}{\sqrt{2}}p_{c,1}^{H}\boldsymbol{h}+\frac{1+i}{\sqrt{2}}p_{c,1}^{V}\boldsymbol{v}\right](\boldsymbol{h}-\boldsymbol{v})$$

$$=\frac{1-i}{\sqrt{2}}(p_{c,1}^{H}+p_{c,2}^{H})(\boldsymbol{hh})+\frac{1-i}{\sqrt{2}}(p_{c,1}^{H}-p_{c,2}^{H})(\boldsymbol{hv})+\frac{1+i}{\sqrt{2}}(p_{c,1}^{V}+p_{c,2}^{V})(\boldsymbol{vh})+\frac{1+i}{\sqrt{2}}(p_{c,1}^{V}-p_{c,2}^{V})(\boldsymbol{vv}).$$

Transformation:

$$\begin{pmatrix}\frac{1-i}{\sqrt{2}}(p_{c,1}^{H}+p_{c,2}^{H})\\ \frac{1-i}{\sqrt{2}}(p_{c,1}^{H}-p_{c,2}^{H})\\ \frac{1+i}{\sqrt{2}}(p_{c,1}^{V}+p_{c,2}^{V})\\ \frac{1+i}{\sqrt{2}}(p_{c,1}^{V}-p_{c,2}^{V})\end{pmatrix}=\begin{pmatrix}\frac{1-i}{\sqrt{2}}&0&0&0\\0&\frac{1-i}{\sqrt{2}}&0&0\\0&0&\frac{1+i}{\sqrt{2}}&0\\0&0&0&\frac{1+i}{\sqrt{2}}\end{pmatrix}\begin{pmatrix}p_{c,1}^{H}+p_{c,2}^{H}\\p_{c,1}^{H}-p_{c,2}^{H}\\p_{c,1}^{V}+p_{c,2}^{V}\\p_{c,1}^{V}-p_{c,2}^{V}\end{pmatrix}=\left(R_z\left(\frac{\pi}{2}\right)\otimes I_2\right)\begin{pmatrix}p_{c,1}^{H}+p_{c,2}^{H}\\p_{c,1}^{H}-p_{c,2}^{H}\\p_{c,1}^{V}+p_{c,2}^{V}\\p_{c,1}^{V}-p_{c,2}^{V}\end{pmatrix}.$$

② Hadamard operation on the second cebit.

Optical setup:

$$\left.\begin{matrix}\left(\frac{1-i}{\sqrt{2}}p_{c,1}^{H}\boldsymbol{h}+\frac{1+i}{\sqrt{2}}p_{c,1}^{V}\boldsymbol{v}\right)f_1\xrightarrow{BD}\left\{\begin{matrix}\frac{1-i}{\sqrt{2}}p_{c,1}^{H}\boldsymbol{h}f_1\\ \frac{1+i}{\sqrt{2}}p_{c,1}^{V}\boldsymbol{v}f_1\end{matrix}\right.\\ \left(\frac{1-i}{\sqrt{2}}p_{c,2}^{H}\boldsymbol{h}+\frac{1+i}{\sqrt{2}}p_{c,2}^{V}\boldsymbol{v}\right)f_2\xrightarrow{BD}\left\{\begin{matrix}\frac{1-i}{\sqrt{2}}p_{c,2}^{H}\boldsymbol{h}f_2\\ \frac{1+i}{\sqrt{2}}p_{c,2}^{V}\boldsymbol{v}f_2\end{matrix}\right.\end{matrix}\right\}\Rightarrow\left\{\begin{matrix}\left.\begin{matrix}\frac{1-i}{\sqrt{2}}p_{c,1}^{H}\boldsymbol{h}f_1\\ \frac{1-i}{\sqrt{2}}p_{c,2}^{H}\boldsymbol{h}f_2\end{matrix}\right\}\xrightarrow{BS}\left\{\begin{matrix}\frac{1-i}{2}(p_{c,1}^{H}+p_{c,2}^{H})\boldsymbol{h}f_1\\ \frac{1-i}{2}(p_{c,1}^{H}-p_{c,2}^{H})\boldsymbol{h}f_2\end{matrix}\right.\\ \left.\begin{matrix}\frac{1+i}{\sqrt{2}}p_{c,1}^{V}\boldsymbol{v}f_1\\ \frac{1+i}{\sqrt{2}}p_{c,2}^{V}\boldsymbol{v}f_2\end{matrix}\right\}\xrightarrow{BS}\left\{\begin{matrix}\frac{1+i}{2}(p_{c,1}^{V}+p_{c,2}^{V})\boldsymbol{v}f_1\\ \frac{1+i}{2}(p_{c,1}^{V}-p_{c,2}^{V})\boldsymbol{v}f_2\end{matrix}\right.\end{matrix}\right..$$

Output state:

$$\left[\frac{1-i}{2}(p_{c,1}^{H}+p_{c,2}^{H})\boldsymbol{h}+\frac{1+i}{2}(p_{c,1}^{V}+p_{c,2}^{V})\boldsymbol{v}\right](\boldsymbol{h}+\boldsymbol{v})$$
$$+\left[\frac{1-i}{2}(p_{c,1}^{H}-p_{c,2}^{H})\boldsymbol{h}+\frac{1+i}{2}(p_{c,1}^{V}-p_{c,2}^{V})\boldsymbol{v}\right](\boldsymbol{h}-\boldsymbol{v})$$
$$=(1-i)p_{c,1}^{H}(\boldsymbol{hh})+(1-i)p_{c,2}^{H}(\boldsymbol{hv})+(1+i)p_{c,1}^{V}(\boldsymbol{vh})+(1+i)p_{c,2}^{V}(\boldsymbol{vv}).$$

Transformation:

$$\begin{pmatrix}(1-i)p_{c,1}^{H}\\(1-i)p_{c,2}^{H}\\(1+i)p_{c,1}^{V}\\(1+i)p_{c,2}^{V}\end{pmatrix}=\begin{pmatrix}\frac{1}{\sqrt{2}}&\frac{1}{\sqrt{2}}&0&0\\ \frac{1}{\sqrt{2}}&\frac{-1}{\sqrt{2}}&0&0\\0&0&\frac{1}{\sqrt{2}}&\frac{1}{\sqrt{2}}\\0&0&\frac{1}{\sqrt{2}}&\frac{-1}{\sqrt{2}}\end{pmatrix}\begin{pmatrix}\frac{1-i}{\sqrt{2}}(p_{c,1}^{H}+p_{c,2}^{H})\\ \frac{1-i}{\sqrt{2}}(p_{c,1}^{H}-p_{c,2}^{H})\\ \frac{1+i}{\sqrt{2}}(p_{c,1}^{V}+p_{c,2}^{V})\\ \frac{1+i}{\sqrt{2}}(p_{c,1}^{V}-p_{c,2}^{V})\end{pmatrix}=(I_2\otimes Ha)\begin{pmatrix}\frac{1-i}{\sqrt{2}}(p_{c,1}^{H}+p_{c,2}^{H})\\ \frac{1-i}{\sqrt{2}}(p_{c,1}^{H}-p_{c,2}^{H})\\ \frac{1+i}{\sqrt{2}}(p_{c,1}^{V}+p_{c,2}^{V})\\ \frac{1+i}{\sqrt{2}}(p_{c,1}^{V}-p_{c,2}^{V})\end{pmatrix}.$$

③ CCX operation.

Optical setup:

$$\left.\begin{matrix}\frac{1-i}{2}\left(p_{c,1}^{H}+p_{c,2}^{H}\right)\boldsymbol{h}f_1\\ \frac{1+i}{2}\left(p_{c,1}^{V}+p_{c,2}^{V}\right)\boldsymbol{v}f_1\end{matrix}\right\}\xrightarrow{PBS}\frac{1-i}{2}\left(p_{c,1}^{H}+p_{c,2}^{H}\right)\boldsymbol{h}f_1+\frac{1+i}{2}\left(p_{c,1}^{V}+p_{c,2}^{V}\right)\boldsymbol{v}f_1,$$

$$\left.\begin{matrix}\frac{1-i}{2}\left(p_{c,1}^{H}-p_{c,2}^{H}\right)\boldsymbol{h}f_2\\ \frac{1+i}{2}\left(p_{c,1}^{V}-p_{c,2}^{V}\right)\boldsymbol{v}f_2\end{matrix}\right\}\xrightarrow{PBS\ with\ a\ \pi-phase\ shift}\frac{1-i}{2}\left(p_{c,1}^{H}-p_{c,2}^{H}\right)\boldsymbol{h}f_2-\frac{1+i}{2}\left(p_{c,1}^{V}-p_{c,2}^{V}\right)\boldsymbol{v}f_2.$$

Output state:

$$\begin{aligned}&\left[\frac{1-i}{2}\left(p_{c,1}^{H}+p_{c,2}^{H}\right)\boldsymbol{h}+\frac{1+i}{2}\left(p_{c,1}^{V}+p_{c,2}^{V}\right)\boldsymbol{v}\right](\boldsymbol{h}+\boldsymbol{v})\\&\qquad+\left[\frac{1-i}{2}\left(p_{c,1}^{H}-p_{c,2}^{H}\right)\boldsymbol{h}-\frac{1+i}{2}\left(p_{c,1}^{V}-p_{c,2}^{V}\right)\boldsymbol{v}\right](\boldsymbol{h}-\boldsymbol{v})\\&=(1-i)p_{c,1}^{H}(\boldsymbol{hh})+(1-i)p_{c,2}^{H}(\boldsymbol{hv})+(1+i)p_{c,2}^{V}(\boldsymbol{vh})+(1+i)p_{c,1}^{V}(\boldsymbol{vv}).\end{aligned}$$

Transformation:

$$\begin{pmatrix}(1-i)p_{c,1}^{H}\\(1-i)p_{c,2}^{H}\\(1+i)p_{c,2}^{V}\\(1+i)p_{c,1}^{V}\end{pmatrix}=\begin{pmatrix}1&0&0&0\\0&1&0&0\\0&0&0&1\\0&0&1&0\end{pmatrix}\begin{pmatrix}(1-i)p_{c,1}^{H}\\(1-i)p_{c,2}^{H}\\(1+i)p_{c,1}^{V}\\(1+i)p_{c,2}^{V}\end{pmatrix}=U_{CCX}\begin{pmatrix}(1-i)p_{c,1}^{H}\\(1-i)p_{c,2}^{H}\\(1+i)p_{c,1}^{V}\\(1+i)p_{c,2}^{V}\end{pmatrix}.$$

④ Hadamard operation on the first cebit.

Optical setup:

$$\begin{aligned}&\frac{1-i}{2}\left(p_{c,1}^{H}+p_{c,2}^{H}\right)\boldsymbol{h}f_1+\frac{1+i}{2}\left(p_{c,1}^{V}+p_{c,2}^{V}\right)\boldsymbol{v}f_1\\&\xrightarrow{HWP@22.5^\circ}\left[\frac{1-i}{2\sqrt{2}}\left(p_{c,1}^{H}+p_{c,2}^{H}\right)+\frac{1+i}{2\sqrt{2}}\left(p_{c,1}^{V}+p_{c,2}^{V}\right)\right]\boldsymbol{h}f_1+\left[\frac{1-i}{2\sqrt{2}}\left(p_{c,1}^{H}+p_{c,2}^{H}\right)-\frac{1+i}{2\sqrt{2}}\left(p_{c,1}^{V}+p_{c,2}^{V}\right)\right]\boldsymbol{v}f_1,\end{aligned}$$

$$\begin{aligned}&\frac{1-i}{2}\left(p_{c,1}^{H}-p_{c,2}^{H}\right)\boldsymbol{h}f_2-\frac{1+i}{2}\left(p_{c,1}^{V}-p_{c,2}^{V}\right)\boldsymbol{v}f_2\\&\xrightarrow{HWP@22.5^\circ}\left[\frac{1-i}{2\sqrt{2}}\left(p_{c,1}^{H}-p_{c,2}^{H}\right)-\frac{1+i}{2\sqrt{2}}\left(p_{c,1}^{V}-p_{c,2}^{V}\right)\right]\boldsymbol{h}f_2+\left[\frac{1-i}{2\sqrt{2}}\left(p_{c,1}^{H}-p_{c,2}^{H}\right)+\frac{1+i}{2\sqrt{2}}\left(p_{c,1}^{V}-p_{c,2}^{V}\right)\right]\boldsymbol{v}f_2.\end{aligned}$$

Output state:

$$\begin{aligned}&\left\{\left[\frac{1-i}{2\sqrt{2}}\left(p_{c,1}^{H}+p_{c,2}^{H}\right)+\frac{1+i}{2\sqrt{2}}\left(p_{c,1}^{V}+p_{c,2}^{V}\right)\right]\boldsymbol{h}+\left[\frac{1-i}{2\sqrt{2}}\left(p_{c,1}^{H}+p_{c,2}^{H}\right)-\frac{1+i}{2\sqrt{2}}\left(p_{c,1}^{V}+p_{c,2}^{V}\right)\right]\boldsymbol{v}\right\}(\boldsymbol{h}+\boldsymbol{v})\\&\qquad+\left\{\left[\frac{1-i}{2\sqrt{2}}\left(p_{c,1}^{H}-p_{c,2}^{H}\right)-\frac{1+i}{2\sqrt{2}}\left(p_{c,1}^{V}-p_{c,2}^{V}\right)\right]\boldsymbol{h}\right.\\&\qquad\left.+\left[\frac{1-i}{2\sqrt{2}}\left(p_{c,1}^{H}-p_{c,2}^{H}\right)+\frac{1+i}{2\sqrt{2}}\left(p_{c,1}^{V}-p_{c,2}^{V}\right)\right]\boldsymbol{v}\right\}(\boldsymbol{h}-\boldsymbol{v})\\&=\left(\frac{1-i}{\sqrt{2}}p_{c,1}^{H}+\frac{1+i}{\sqrt{2}}p_{c,2}^{V}\right)(\boldsymbol{hh})+\left(\frac{1-i}{\sqrt{2}}p_{c,2}^{H}+\frac{1+i}{\sqrt{2}}p_{c,1}^{V}\right)(\boldsymbol{hv})+\left(\frac{1-i}{\sqrt{2}}p_{c,1}^{H}-\frac{1+i}{\sqrt{2}}p_{c,2}^{V}\right)(\boldsymbol{vh})\\&\qquad+\left(\frac{1-i}{\sqrt{2}}p_{c,2}^{H}-\frac{1+i}{\sqrt{2}}p_{c,1}^{V}\right)(\boldsymbol{vv}).\end{aligned}$$

Transformation:

$$\begin{pmatrix} \frac{1-i}{\sqrt{2}}p_{c,1}^{H}+\frac{1+i}{\sqrt{2}}p_{c,2}^{V} \\ \frac{1-i}{\sqrt{2}}p_{c,2}^{H}+\frac{1+i}{\sqrt{2}}p_{c,1}^{V} \\ \frac{1-i}{\sqrt{2}}p_{c,1}^{H}-\frac{1+i}{\sqrt{2}}p_{c,2}^{V} \\ \frac{1-i}{\sqrt{2}}p_{c,2}^{H}-\frac{1+i}{\sqrt{2}}p_{c,1}^{V} \end{pmatrix} = \begin{pmatrix} \frac{1}{\sqrt{2}} & 0 & \frac{1}{\sqrt{2}} & 0 \\ 0 & \frac{1}{\sqrt{2}} & 0 & \frac{1}{\sqrt{2}} \\ \frac{1}{\sqrt{2}} & 0 & -\frac{1}{\sqrt{2}} & 0 \\ 0 & \frac{1}{\sqrt{2}} & 0 & -\frac{1}{\sqrt{2}} \end{pmatrix} \begin{pmatrix} (1-i)p_{c,1}^{H} \\ (1-i)p_{c,2}^{H} \\ (1+i)p_{c,2}^{V} \\ (1+i)p_{c,1}^{V} \end{pmatrix} = (Ha\otimes I_2)\begin{pmatrix} (1-i)p_{c,1}^{H} \\ (1-i)p_{c,2}^{H} \\ (1+i)p_{c,2}^{V} \\ (1+i)p_{c,1}^{V} \end{pmatrix}.$$

⑤ $R_Y(\pi/2)$ on the second cebit.

Optical setup:

$$\left[\frac{1-i}{2\sqrt{2}}\left(p_{c,1}^{H}+p_{c,2}^{H}\right)+\frac{1+i}{2\sqrt{2}}\left(p_{c,1}^{V}+p_{c,2}^{V}\right)\right]\boldsymbol{h}f_1+\left[\frac{1-i}{2\sqrt{2}}\left(p_{c,1}^{H}+p_{c,2}^{H}\right)-\frac{1+i}{2\sqrt{2}}\left(p_{c,1}^{V}+p_{c,2}^{V}\right)\right]\boldsymbol{v}f_1$$

$$\xrightarrow{BD}\begin{cases}\left[\frac{1-i}{2\sqrt{2}}\left(p_{c,1}^{H}+p_{c,2}^{H}\right)+\frac{1+i}{2\sqrt{2}}\left(p_{c,1}^{V}+p_{c,2}^{V}\right)\right]\boldsymbol{h}f_1 \\ \left[\frac{1-i}{2\sqrt{2}}\left(p_{c,1}^{H}+p_{c,2}^{H}\right)-\frac{1+i}{2\sqrt{2}}\left(p_{c,1}^{V}+p_{c,2}^{V}\right)\right]\boldsymbol{v}f_1\end{cases}, \qquad \text{(mode 1)}$$

$$\left[\frac{1-i}{2\sqrt{2}}\left(p_{c,1}^{H}-p_{c,2}^{H}\right)-\frac{1+i}{2\sqrt{2}}\left(p_{c,1}^{V}-p_{c,2}^{V}\right)\right]\boldsymbol{h}f_2+\left[\frac{1-i}{2\sqrt{2}}\left(p_{c,1}^{H}-p_{c,2}^{H}\right)+\frac{1+i}{2\sqrt{2}}\left(p_{c,1}^{V}-p_{c,2}^{V}\right)\right]\boldsymbol{v}f_2$$

$$\xrightarrow{BD}\begin{cases}\left[\frac{1-i}{2\sqrt{2}}\left(p_{c,1}^{H}-p_{c,2}^{H}\right)-\frac{1+i}{2\sqrt{2}}\left(p_{c,1}^{V}-p_{c,2}^{V}\right)\right]\boldsymbol{h}f_2 \\ \left[\frac{1-i}{2\sqrt{2}}\left(p_{c,1}^{H}-p_{c,2}^{H}\right)+\frac{1+i}{2\sqrt{2}}\left(p_{c,1}^{V}-p_{c,2}^{V}\right)\right]\boldsymbol{v}f_2\end{cases}, \qquad \text{(mode 2)}$$

$$\left.\begin{matrix}\text{(mode 1)}\\ \text{(mode 2)}\end{matrix}\right\}$$

$$\Rightarrow\begin{cases}\left.\begin{matrix}\left[\frac{1-i}{2\sqrt{2}}\left(p_{c,1}^{H}+p_{c,2}^{H}\right)+\frac{1+i}{2\sqrt{2}}\left(p_{c,1}^{V}+p_{c,2}^{V}\right)\right]\boldsymbol{h}f_1 \\ \left[\frac{1-i}{2\sqrt{2}}\left(p_{c,1}^{H}-p_{c,2}^{H}\right)-\frac{1+i}{2\sqrt{2}}\left(p_{c,1}^{V}-p_{c,2}^{V}\right)\right]\boldsymbol{h}f_2\end{matrix}\right\}\xrightarrow{BS\ with\ a\ \pi-phase\ shift}\begin{cases}\left(\frac{1-i}{2}p_{c,1}^{H}+\frac{1+i}{2}p_{c,2}^{V}\right)\boldsymbol{h}f_1 \\ -\left(\frac{1-i}{2}p_{c,2}^{H}+\frac{1+i}{2}p_{c,1}^{V}\right)\boldsymbol{h}f_2\end{cases} \\ \left.\begin{matrix}\left[\frac{1-i}{2\sqrt{2}}\left(p_{c,1}^{H}+p_{c,2}^{H}\right)-\frac{1+i}{2\sqrt{2}}\left(p_{c,1}^{V}+p_{c,2}^{V}\right)\right]\boldsymbol{v}f_1 \\ \left[\frac{1-i}{2\sqrt{2}}\left(p_{c,1}^{H}-p_{c,2}^{H}\right)+\frac{1+i}{2\sqrt{2}}\left(p_{c,1}^{V}-p_{c,2}^{V}\right)\right]\boldsymbol{v}f_2\end{matrix}\right\}\xrightarrow{BS\ with\ a\ \pi-phase\ shift}\begin{cases}\left(\frac{1-i}{2}p_{c,1}^{H}-\frac{1+i}{2}p_{c,2}^{V}\right)\boldsymbol{v}f_1 \\ -\left(\frac{1-i}{2}p_{c,2}^{H}-\frac{1+i}{2}p_{c,1}^{V}\right)\boldsymbol{v}f_2\end{cases}\end{cases},$$

Output state:

$$\left[\left(\frac{1-i}{2}p_{c,1}^{H}+\frac{1+i}{2}p_{c,2}^{V}\right)\boldsymbol{h}+\left(\frac{1-i}{2}p_{c,1}^{H}-\frac{1+i}{2}p_{c,2}^{V}\right)\boldsymbol{v}\right](\boldsymbol{h}+\boldsymbol{v})$$
$$-\left[\left(\frac{1-i}{2}p_{c,2}^{H}+\frac{1+i}{2}p_{c,1}^{V}\right)\boldsymbol{h}+\left(\frac{1-i}{2}p_{c,2}^{H}-\frac{1+i}{2}p_{c,1}^{V}\right)\boldsymbol{v}\right](\boldsymbol{h}-\boldsymbol{v})$$

$$= \left[\frac{1-i}{2}\left(p_{c,1}^{H} - p_{c,2}^{H}\right) + \frac{1+i}{2}\left(p_{c,2}^{V} - p_{c,1}^{V}\right)\right](\boldsymbol{hh}) + \left[\frac{1-i}{2}\left(p_{c,1}^{H} + p_{c,2}^{H}\right) + \frac{1+i}{2}\left(p_{c,2}^{V} + p_{c,1}^{V}\right)\right](\boldsymbol{hv})$$
$$+ \left[\frac{1-i}{2}\left(p_{c,1}^{H} - p_{c,2}^{H}\right) - \frac{1+i}{2}\left(p_{c,2}^{V} - p_{c,1}^{V}\right)\right](\boldsymbol{vh})$$
$$+ \left[\frac{1-i}{2}\left(p_{c,1}^{H} + p_{c,2}^{H}\right) - \frac{1+i}{2}\left(p_{c,2}^{V} + p_{c,1}^{V}\right)\right](\boldsymbol{vv}).$$

Transformation:

$$\begin{pmatrix} \frac{1-i}{2}\left(p_{c,1}^{H} - p_{c,2}^{H}\right) + \frac{1+i}{2}\left(p_{c,2}^{V} - p_{c,1}^{V}\right) \\ \frac{1-i}{2}\left(p_{c,1}^{H} + p_{c,2}^{H}\right) + \frac{1+i}{2}\left(p_{c,2}^{V} + p_{c,1}^{V}\right) \\ \frac{1-i}{2}\left(p_{c,1}^{H} - p_{c,2}^{H}\right) - \frac{1+i}{2}\left(p_{c,2}^{V} - p_{c,1}^{V}\right) \\ \frac{1-i}{2}\left(p_{c,1}^{H} + p_{c,2}^{H}\right) - \frac{1+i}{2}\left(p_{c,2}^{V} + p_{c,1}^{V}\right) \end{pmatrix} = \begin{pmatrix} \frac{1}{\sqrt{2}} & -\frac{1}{\sqrt{2}} & 0 & 0 \\ \frac{1}{\sqrt{2}} & \frac{1}{\sqrt{2}} & 0 & 0 \\ 0 & 0 & \frac{1}{\sqrt{2}} & -\frac{1}{\sqrt{2}} \\ 0 & 0 & \frac{1}{\sqrt{2}} & \frac{1}{\sqrt{2}} \end{pmatrix} \begin{pmatrix} \frac{1-i}{\sqrt{2}}p_{c,1}^{H} + \frac{1+i}{\sqrt{2}}p_{c,2}^{V} \\ \frac{1-i}{\sqrt{2}}p_{c,2}^{H} + \frac{1+i}{\sqrt{2}}p_{c,1}^{V} \\ \frac{1-i}{\sqrt{2}}p_{c,1}^{H} - \frac{1+i}{\sqrt{2}}p_{c,2}^{V} \\ \frac{1-i}{\sqrt{2}}p_{c,2}^{H} - \frac{1+i}{\sqrt{2}}p_{c,1}^{V} \end{pmatrix}$$

$$= \left(I_2 \otimes R_Y\left(\frac{\pi}{2}\right)\right) \begin{pmatrix} \frac{1-i}{\sqrt{2}}p_{c,1}^{H} + \frac{1+i}{\sqrt{2}}p_{c,2}^{V} \\ \frac{1-i}{\sqrt{2}}p_{c,2}^{H} + \frac{1+i}{\sqrt{2}}p_{c,1}^{V} \\ \frac{1-i}{\sqrt{2}}p_{c,1}^{H} - \frac{1+i}{\sqrt{2}}p_{c,2}^{V} \\ \frac{1-i}{\sqrt{2}}p_{c,2}^{H} - \frac{1+i}{\sqrt{2}}p_{c,1}^{V} \end{pmatrix}.$$

⑥ CCX operation.

Optical setup:

$$\left.\begin{matrix} \left(\frac{1-i}{2}p_{c,1}^{H} + \frac{1+i}{2}p_{c,2}^{V}\right)\boldsymbol{h}f_1 \\ \left(\frac{1-i}{2}p_{c,1}^{H} - \frac{1+i}{2}p_{c,2}^{V}\right)\boldsymbol{v}f_1 \end{matrix}\right\} \xrightarrow{PBS} \left(\frac{1-i}{2}p_{c,1}^{H} + \frac{1+i}{2}p_{c,2}^{V}\right)\boldsymbol{h}f_1 + \left(\frac{1-i}{2}p_{c,1}^{H} - \frac{1+i}{2}p_{c,2}^{V}\right)\boldsymbol{v}f_1,$$

$$\left.\begin{matrix} -\left(\frac{1-i}{2}p_{c,2}^{H} + \frac{1+i}{2}p_{c,1}^{V}\right)\boldsymbol{h}f_2 \\ -\left(\frac{1-i}{2}p_{c,2}^{H} - \frac{1+i}{2}p_{c,1}^{V}\right)\boldsymbol{v}f_2 \end{matrix}\right\} \xrightarrow{PBS\ with\ a\ \pi\text{-}phase\ shift} -\left(\frac{1-i}{2}p_{c,2}^{H} + \frac{1+i}{2}p_{c,1}^{V}\right)\boldsymbol{h}f_2$$
$$+ \left(\frac{1-i}{2}p_{c,2}^{H} - \frac{1+i}{2}p_{c,1}^{V}\right)\boldsymbol{v}f_2.$$

Output state:

$$\left[\left(\frac{1-i}{2}p_{c,1}^{H} + \frac{1+i}{2}p_{c,2}^{V}\right)\boldsymbol{h} + \left(\frac{1-i}{2}p_{c,1}^{H} - \frac{1+i}{2}p_{c,2}^{V}\right)\boldsymbol{v}\right](\boldsymbol{h} + \boldsymbol{v})$$
$$- \left[\left(\frac{1-i}{2}p_{c,2}^{H} + \frac{1+i}{2}p_{c,1}^{V}\right)\boldsymbol{h} - \left(\frac{1-i}{2}p_{c,2}^{H} - \frac{1+i}{2}p_{c,1}^{V}\right)\boldsymbol{v}\right](\boldsymbol{h} - \boldsymbol{v})$$

$$= \left[\frac{1-i}{2}\left(p_{c,1}^H - p_{c,2}^H\right) + \frac{1+i}{2}\left(p_{c,2}^V - p_{c,1}^V\right)\right](\boldsymbol{hh}) + \left[\frac{1-i}{2}\left(p_{c,1}^H + p_{c,2}^H\right) + \frac{1+i}{2}\left(p_{c,2}^V + p_{c,1}^V\right)\right](\boldsymbol{hv})$$
$$+ \left[\frac{1-i}{2}\left(p_{c,1}^H + p_{c,2}^H\right) - \frac{1+i}{2}\left(p_{c,2}^V + p_{c,1}^V\right)\right](\boldsymbol{vh})$$
$$+ \left[\frac{1-i}{2}\left(p_{c,1}^H - p_{c,2}^H\right) - \frac{1+i}{2}\left(p_{c,2}^V - p_{c,1}^V\right)\right](\boldsymbol{vv}).$$

Transformation:

$$\begin{pmatrix} \frac{1-i}{2}\left(p_{c,1}^H - p_{c,2}^H\right) + \frac{1+i}{2}\left(p_{c,2}^V - p_{c,1}^V\right) \\ \frac{1-i}{2}\left(p_{c,1}^H + p_{c,2}^H\right) + \frac{1+i}{2}\left(p_{c,2}^V + p_{c,1}^V\right) \\ \frac{1-i}{2}\left(p_{c,1}^H + p_{c,2}^H\right) - \frac{1+i}{2}\left(p_{c,2}^V + p_{c,1}^V\right) \\ \frac{1-i}{2}\left(p_{c,1}^H - p_{c,2}^H\right) - \frac{1+i}{2}\left(p_{c,2}^V - p_{c,1}^V\right) \end{pmatrix} = \begin{pmatrix} 1 & 0 & 0 & 0 \\ 0 & 1 & 0 & 0 \\ 0 & 0 & 0 & 1 \\ 0 & 0 & 1 & 0 \end{pmatrix} \begin{pmatrix} \frac{1-i}{2}\left(p_{c,1}^H - p_{c,2}^H\right) + \frac{1+i}{2}\left(p_{c,2}^V - p_{c,1}^V\right) \\ \frac{1-i}{2}\left(p_{c,1}^H + p_{c,2}^H\right) + \frac{1+i}{2}\left(p_{c,2}^V + p_{c,1}^V\right) \\ \frac{1-i}{2}\left(p_{c,1}^H - p_{c,2}^H\right) - \frac{1+i}{2}\left(p_{c,2}^V - p_{c,1}^V\right) \\ \frac{1-i}{2}\left(p_{c,1}^H + p_{c,2}^H\right) - \frac{1+i}{2}\left(p_{c,2}^V + p_{c,1}^V\right) \end{pmatrix}$$
$$= U_{CCX} \begin{pmatrix} \frac{1-i}{2}\left(p_{c,1}^H - p_{c,2}^H\right) + \frac{1+i}{2}\left(p_{c,2}^V - p_{c,1}^V\right) \\ \frac{1-i}{2}\left(p_{c,1}^H + p_{c,2}^H\right) + \frac{1+i}{2}\left(p_{c,2}^V + p_{c,1}^V\right) \\ \frac{1-i}{2}\left(p_{c,1}^H - p_{c,2}^H\right) - \frac{1+i}{2}\left(p_{c,2}^V - p_{c,1}^V\right) \\ \frac{1-i}{2}\left(p_{c,1}^H + p_{c,2}^H\right) - \frac{1+i}{2}\left(p_{c,2}^V + p_{c,1}^V\right) \end{pmatrix}.$$

⑦ $R_X(\pi/2)$ on the first cebit.

Optical setup:

$$\left(\frac{1-i}{2}p_{c,1}^H + \frac{1+i}{2}p_{c,2}^V\right)\boldsymbol{h}f_1 + \left(\frac{1-i}{2}p_{c,1}^H - \frac{1+i}{2}p_{c,2}^V\right)\boldsymbol{v}f_1 \xrightarrow{QWP@45^\circ} \left(\frac{-i}{\sqrt{2}}p_{c,1}^H + \frac{i}{\sqrt{2}}p_{c,2}^V\right)\boldsymbol{h}f_1$$
$$+ \left(\frac{-i}{\sqrt{2}}p_{c,1}^H - \frac{i}{\sqrt{2}}p_{c,2}^V\right)\boldsymbol{v}f_1,$$
$$-\left(\frac{1-i}{2}p_{c,2}^H + \frac{1+i}{2}p_{c,1}^V\right)\boldsymbol{h}f_2 + \left(\frac{1-i}{2}p_{c,2}^H - \frac{1+i}{2}p_{c,1}^V\right)\boldsymbol{v}f_2 \xrightarrow{QWP@45^\circ} -\left(\frac{1}{\sqrt{2}}p_{c,2}^H + \frac{1}{\sqrt{2}}p_{c,1}^V\right)\boldsymbol{h}f_1$$
$$+ \left(\frac{1}{\sqrt{2}}p_{c,2}^H - \frac{1}{\sqrt{2}}p_{c,1}^V\right)\boldsymbol{v}f_1.$$

Output state:

$$\left[\left(\frac{-i}{\sqrt{2}}p_{c,1}^H + \frac{i}{\sqrt{2}}p_{c,2}^V\right)\boldsymbol{h} + \left(\frac{-i}{\sqrt{2}}p_{c,1}^H - \frac{i}{\sqrt{2}}p_{c,2}^V\right)\boldsymbol{v}\right](\boldsymbol{h}+\boldsymbol{v})$$
$$- \left[\left(\frac{1}{\sqrt{2}}p_{c,2}^H + \frac{1}{\sqrt{2}}p_{c,1}^V\right)\boldsymbol{h} - \left(\frac{1}{\sqrt{2}}p_{c,2}^H - \frac{1}{\sqrt{2}}p_{c,1}^V\right)\boldsymbol{v}\right](\boldsymbol{h}-\boldsymbol{v})$$
$$= -\left(\frac{i}{\sqrt{2}}p_{c,1}^H + \frac{1}{\sqrt{2}}p_{c,2}^H + \frac{1}{\sqrt{2}}p_{c,1}^V - \frac{i}{\sqrt{2}}p_{c,2}^V\right)(\boldsymbol{hh}) - \left(\frac{i}{\sqrt{2}}p_{c,1}^H - \frac{1}{\sqrt{2}}p_{c,2}^H - \frac{1}{\sqrt{2}}p_{c,1}^V - \frac{i}{\sqrt{2}}p_{c,2}^V\right)(\boldsymbol{hv})$$
$$- \left(\frac{i}{\sqrt{2}}p_{c,1}^H - \frac{1}{\sqrt{2}}p_{c,2}^H + \frac{1}{\sqrt{2}}p_{c,1}^V + \frac{i}{\sqrt{2}}p_{c,2}^V\right)(\boldsymbol{vh})$$
$$- \left(\frac{i}{\sqrt{2}}p_{c,1}^H + \frac{1}{\sqrt{2}}p_{c,2}^H - \frac{1}{\sqrt{2}}p_{c,1}^V + \frac{i}{\sqrt{2}}p_{c,2}^V\right)(\boldsymbol{vv}).$$

Transformation:

$$\begin{pmatrix} -\left(\frac{i}{\sqrt{2}}p_{c,1}^{H}+\frac{1}{\sqrt{2}}p_{c,2}^{H}+\frac{1}{\sqrt{2}}p_{c,1}^{V}-\frac{i}{\sqrt{2}}p_{c,2}^{V}\right) \\ -\left(\frac{i}{\sqrt{2}}p_{c,1}^{H}-\frac{1}{\sqrt{2}}p_{c,2}^{H}-\frac{1}{\sqrt{2}}p_{c,1}^{V}-\frac{i}{\sqrt{2}}p_{c,2}^{V}\right) \\ -\left(\frac{i}{\sqrt{2}}p_{c,1}^{H}-\frac{1}{\sqrt{2}}p_{c,2}^{H}+\frac{1}{\sqrt{2}}p_{c,1}^{V}+\frac{i}{\sqrt{2}}p_{c,2}^{V}\right) \\ -\left(\frac{i}{\sqrt{2}}p_{c,1}^{H}+\frac{1}{\sqrt{2}}p_{c,2}^{H}-\frac{1}{\sqrt{2}}p_{c,1}^{V}+\frac{i}{\sqrt{2}}p_{c,2}^{V}\right) \end{pmatrix}$$

$$= \begin{pmatrix} \frac{1}{\sqrt{2}} & 0 & -\frac{i}{\sqrt{2}} & 0 \\ 0 & \frac{1}{\sqrt{2}} & 0 & -\frac{i}{\sqrt{2}} \\ -\frac{i}{\sqrt{2}} & 0 & \frac{1}{\sqrt{2}} & 0 \\ 0 & -\frac{i}{\sqrt{2}} & 0 & \frac{1}{\sqrt{2}} \end{pmatrix} \begin{pmatrix} \frac{1-i}{2}\left(p_{c,1}^{H}-p_{c,2}^{H}\right)+\frac{1+i}{2}\left(p_{c,2}^{V}-p_{c,1}^{V}\right) \\ \frac{1-i}{2}\left(p_{c,1}^{H}+p_{c,2}^{H}\right)+\frac{1+i}{2}\left(p_{c,2}^{V}+p_{c,1}^{V}\right) \\ \frac{1-i}{2}\left(p_{c,1}^{H}+p_{c,2}^{H}\right)-\frac{1+i}{2}\left(p_{c,2}^{V}+p_{c,1}^{V}\right) \\ \frac{1-i}{2}\left(p_{c,1}^{H}-p_{c,2}^{H}\right)-\frac{1+i}{2}\left(p_{c,2}^{V}-p_{c,1}^{V}\right) \end{pmatrix}$$

$$= \left(R_X\left(\frac{\pi}{2}\right)\otimes I_2\right) \begin{pmatrix} \frac{1-i}{2}\left(p_{c,1}^{H}-p_{c,2}^{H}\right)+\frac{1+i}{2}\left(p_{c,2}^{V}-p_{c,1}^{V}\right) \\ \frac{1-i}{2}\left(p_{c,1}^{H}+p_{c,2}^{H}\right)+\frac{1+i}{2}\left(p_{c,2}^{V}+p_{c,1}^{V}\right) \\ \frac{1-i}{2}\left(p_{c,1}^{H}+p_{c,2}^{H}\right)-\frac{1+i}{2}\left(p_{c,2}^{V}+p_{c,1}^{V}\right) \\ \frac{1-i}{2}\left(p_{c,1}^{H}-p_{c,2}^{H}\right)-\frac{1+i}{2}\left(p_{c,2}^{V}-p_{c,1}^{V}\right) \end{pmatrix}.$$

⑧ CCX operation.

Optical setup:

$$\left(\frac{-i}{\sqrt{2}}p_{c,1}^{H}+\frac{i}{\sqrt{2}}p_{c,2}^{V}\right)\boldsymbol{h}f_1+\left(\frac{-i}{\sqrt{2}}p_{c,1}^{H}-\frac{i}{\sqrt{2}}p_{c,2}^{V}\right)\boldsymbol{v}f_1 \rightarrow \left(\frac{-i}{\sqrt{2}}p_{c,1}^{H}+\frac{i}{\sqrt{2}}p_{c,2}^{V}\right)\boldsymbol{h}f_1 + \left(\frac{-i}{\sqrt{2}}p_{c,1}^{H}-\frac{i}{\sqrt{2}}p_{c,2}^{V}\right)\boldsymbol{v}f_1,$$

$$-\left(\frac{1}{\sqrt{2}}p_{c,2}^{H}+\frac{1}{\sqrt{2}}p_{c,1}^{V}\right)\boldsymbol{h}f_1+\left(\frac{1}{\sqrt{2}}p_{c,2}^{H}-\frac{1}{\sqrt{2}}p_{c,1}^{V}\right)\boldsymbol{v}f_1 \xrightarrow{HWP@0^\circ} -\left(\frac{1}{\sqrt{2}}p_{c,2}^{H}+\frac{1}{\sqrt{2}}p_{c,1}^{V}\right)\boldsymbol{h}f_1 - \left(\frac{1}{\sqrt{2}}p_{c,2}^{H}-\frac{1}{\sqrt{2}}p_{c,1}^{V}\right)\boldsymbol{v}f_1.$$

Output state:

$$\left[\left(\frac{-i}{\sqrt{2}}p_{c,1}^{H}+\frac{i}{\sqrt{2}}p_{c,2}^{V}\right)\boldsymbol{h}+\left(\frac{-i}{\sqrt{2}}p_{c,1}^{H}-\frac{i}{\sqrt{2}}p_{c,2}^{V}\right)\boldsymbol{v}\right](\boldsymbol{h}+\boldsymbol{v}) - \left[\left(\frac{1}{\sqrt{2}}p_{c,2}^{H}+\frac{1}{\sqrt{2}}p_{c,1}^{V}\right)\boldsymbol{h}+\left(\frac{1}{\sqrt{2}}p_{c,2}^{H}-\frac{1}{\sqrt{2}}p_{c,1}^{V}\right)\boldsymbol{v}\right](\boldsymbol{h}-\boldsymbol{v})$$

$$= -\left(\frac{i}{\sqrt{2}}p_{c,1}^{H} + \frac{1}{\sqrt{2}}p_{c,2}^{H} + \frac{1}{\sqrt{2}}p_{c,1}^{V} - \frac{i}{\sqrt{2}}p_{c,2}^{V}\right)(\boldsymbol{hh}) - \left(\frac{i}{\sqrt{2}}p_{c,1}^{H} - \frac{1}{\sqrt{2}}p_{c,2}^{H} - \frac{1}{\sqrt{2}}p_{c,1}^{V} - \frac{i}{\sqrt{2}}p_{c,2}^{V}\right)(\boldsymbol{hv})$$
$$- \left(\frac{i}{\sqrt{2}}p_{c,1}^{H} + \frac{1}{\sqrt{2}}p_{c,2}^{H} - \frac{1}{\sqrt{2}}p_{c,1}^{V} + \frac{i}{\sqrt{2}}p_{c,2}^{V}\right)(\boldsymbol{vh})$$
$$- \left(\frac{i}{\sqrt{2}}p_{c,1}^{H} - \frac{1}{\sqrt{2}}p_{c,2}^{H} + \frac{1}{\sqrt{2}}p_{c,1}^{V} + \frac{i}{\sqrt{2}}p_{c,2}^{V}\right)(\boldsymbol{vv}).$$

Transformation:

$$\begin{pmatrix} -\left(\frac{i}{\sqrt{2}}p_{c,1}^{H} + \frac{1}{\sqrt{2}}p_{c,2}^{H} + \frac{1}{\sqrt{2}}p_{c,1}^{V} - \frac{i}{\sqrt{2}}p_{c,2}^{V}\right) \\ -\left(\frac{i}{\sqrt{2}}p_{c,1}^{H} - \frac{1}{\sqrt{2}}p_{c,2}^{H} - \frac{1}{\sqrt{2}}p_{c,1}^{V} - \frac{i}{\sqrt{2}}p_{c,2}^{V}\right) \\ -\left(\frac{i}{\sqrt{2}}p_{c,1}^{H} + \frac{1}{\sqrt{2}}p_{c,2}^{H} - \frac{1}{\sqrt{2}}p_{c,1}^{V} + \frac{i}{\sqrt{2}}p_{c,2}^{V}\right) \\ -\left(\frac{i}{\sqrt{2}}p_{c,1}^{H} - \frac{1}{\sqrt{2}}p_{c,2}^{H} + \frac{1}{\sqrt{2}}p_{c,1}^{V} + \frac{i}{\sqrt{2}}p_{c,2}^{V}\right) \end{pmatrix}$$

$$= \begin{pmatrix} 1 & 0 & 0 & 0 \\ 0 & 1 & 0 & 0 \\ 0 & 0 & 0 & 1 \\ 0 & 0 & 1 & 0 \end{pmatrix} \begin{pmatrix} -\left(\frac{i}{\sqrt{2}}p_{c,1}^{H} + \frac{1}{\sqrt{2}}p_{c,2}^{H} + \frac{1}{\sqrt{2}}p_{c,1}^{V} - \frac{i}{\sqrt{2}}p_{c,2}^{V}\right) \\ -\left(\frac{i}{\sqrt{2}}p_{c,1}^{H} - \frac{1}{\sqrt{2}}p_{c,2}^{H} - \frac{1}{\sqrt{2}}p_{c,1}^{V} - \frac{i}{\sqrt{2}}p_{c,2}^{V}\right) \\ -\left(\frac{i}{\sqrt{2}}p_{c,1}^{H} - \frac{1}{\sqrt{2}}p_{c,2}^{H} + \frac{1}{\sqrt{2}}p_{c,1}^{V} + \frac{i}{\sqrt{2}}p_{c,2}^{V}\right) \\ -\left(\frac{i}{\sqrt{2}}p_{c,1}^{H} + \frac{1}{\sqrt{2}}p_{c,2}^{H} - \frac{1}{\sqrt{2}}p_{c,1}^{V} + \frac{i}{\sqrt{2}}p_{c,2}^{V}\right) \end{pmatrix}$$

$$= U_{CCX} \begin{pmatrix} -\left(\frac{i}{\sqrt{2}}p_{c,1}^{H} + \frac{1}{\sqrt{2}}p_{c,2}^{H} + \frac{1}{\sqrt{2}}p_{c,1}^{V} - \frac{i}{\sqrt{2}}p_{c,2}^{V}\right) \\ -\left(\frac{i}{\sqrt{2}}p_{c,1}^{H} - \frac{1}{\sqrt{2}}p_{c,2}^{H} - \frac{1}{\sqrt{2}}p_{c,1}^{V} - \frac{i}{\sqrt{2}}p_{c,2}^{V}\right) \\ -\left(\frac{i}{\sqrt{2}}p_{c,1}^{H} - \frac{1}{\sqrt{2}}p_{c,2}^{H} + \frac{1}{\sqrt{2}}p_{c,1}^{V} + \frac{i}{\sqrt{2}}p_{c,2}^{V}\right) \\ -\left(\frac{i}{\sqrt{2}}p_{c,1}^{H} + \frac{1}{\sqrt{2}}p_{c,2}^{H} - \frac{1}{\sqrt{2}}p_{c,1}^{V} + \frac{i}{\sqrt{2}}p_{c,2}^{V}\right) \end{pmatrix}.$$

Therefore, the total operation can be given by

$$U_{CCX}\left(R_X\left(\frac{\pi}{2}\right) \otimes I_2\right) U_{CCX}(Ha \otimes I_2)\left(I_2 \otimes R_Y\left(\frac{\pi}{2}\right)\ \right) U_{CCX}\left(R_z\left(\frac{\pi}{2}\right) \otimes I_2\right)(I_2 \otimes Ha)$$
$$= U_{CCX}\left(R_X\left(\frac{\pi}{2}\right) \otimes I_2\right) U_{CCX}\left(Ha \otimes R_Y\left(\frac{\pi}{2}\right)\right) U_{CCX}\left(R_z\left(\frac{\pi}{2}\right) \otimes Ha\right)$$
$$= \frac{\sqrt{2}}{4}\begin{pmatrix} -1-i & 1-i & -1+i & -1-i \\ 1-i & -1-i & 1+i & 1-i \\ -1-i & 1-i & 1-i & 1+i \\ 1-i & -1-i & -1-i & -1+i \end{pmatrix}. \tag{s26}$$

According to the process tomography theory and the Kraus theorem [9], a 4-by-4 unitary operator can be expressed by a sum of the Kronecker products of the operators $\{I_2, X, -iY, Z\}$, whose coefficients form a 16-by-16 matrix denoted by $\chi$. The $\chi$ matrix of (s26) is given by figure 5 in the main text.

## S4: A brief discussion on the beam generation

The multi-mode polarized beam we consider in the main text (equation (2)) can be generalized in many ways. Also, for some special cases, it can be generated with a low-cost setup. We present a simple example here. Suppose that $f_k$ is a bunch of instinct pules, satisfying the orthonormal relation $\int f_{k_1}(\boldsymbol{r},t) \dots f_{k_P}(\boldsymbol{r},t) d\Omega = \delta_{k_1,\dots,k_P}$. Then, a beam composed of those modes can be generated in the following way. We define a general pulse function by $g(t,t_0)$, where $t$ is the time variable and $t_0$ is the center location of the pulse. The width of the pulse is denoted by $w$. Firstly, a pulse field $g(t,t_0)$ can be generated by a pulse laser. Then, input the pulse to an interferometer, the one arm of which has a $\Delta$ delayer such that $g(t,t_0) \rightarrow g(t,t_0+\Delta)$, $\Delta > w$. Next, using a similar interferometer composed of a $2\Delta$ delayer, one can double the pulse number. A schematic graph is shown by figure S6, and the pulse output by the interferometers can be expressed by

$$g(t,t_0) \overset{\text{(i)}}{\rightarrow} \begin{Bmatrix} g(t,t_0) \\ g(t,t_0+\Delta) \end{Bmatrix} \overset{\text{(ii)}}{\rightarrow} \begin{Bmatrix} g(t,t_0)+g(t,t_0+\Delta) \\ g(t,t_0+2\Delta)-g(t,t_0+3\Delta) \end{Bmatrix}$$
$$\overset{\text{(iii)}}{\longrightarrow} \begin{Bmatrix} g(t,t_0)+g(t,t_0+\Delta)+g(t,t_0+2\Delta)-g(t,t_0+3\Delta) \\ g(t,t_0+4\Delta)+g(t,t_0+5\Delta)-g(t,t_0+6\Delta)+g(t,t_0+7\Delta) \end{Bmatrix} \rightarrow \cdots$$
$$\overset{\text{(iv)}}{\longrightarrow} \begin{Bmatrix} g(t,t_0) \dots g(t,t_0+(2^{N-1}-1)\Delta) \\ g(t,t_0+2^{N-1}\Delta) \dots g(t,t_0+(2^N-1)\Delta) \end{Bmatrix} \xrightarrow{\text{output}} g(t,t_0), g(t,t_0+\Delta), \dots, g(t,t_0+(2^N-1)\Delta) \quad \text{(s26)}$$

Finally, after the above $N$ interferometers with $N$ delayers, one can generate $2^N$ pulses. The phases of the pulses are sometimes opposite, which is not relevant and can be adjusted to the same easily. Also, the intensities of the pulses are largely decreased. This does not affect the orthonormality of the pulses, which is the basement of the optical computing scheme. Besides, for classical light fields, the output intensity can be enhanced by light amplifiers, or simply applying a laser with higher power. Of course, such a method is not suitable for encoding all cebit states. However, it may be beneficial in certain tasks, such as the simulation of advantageous quantum algorithms.

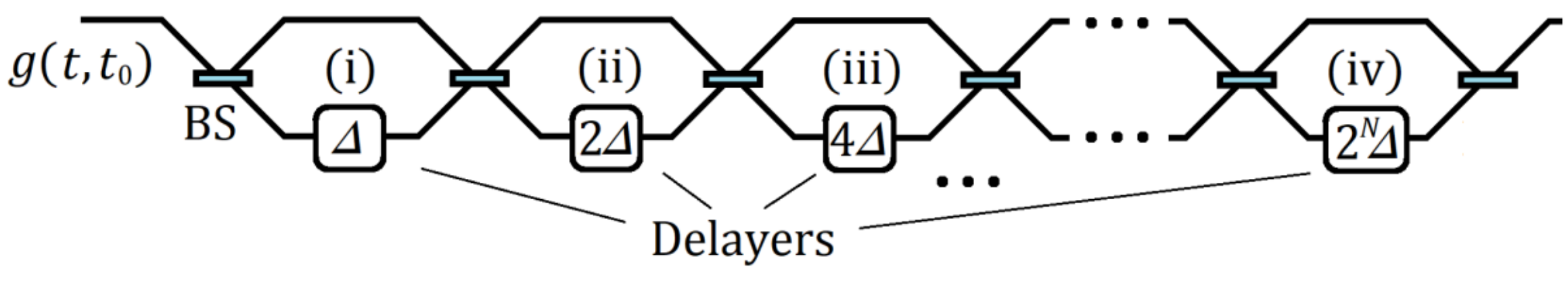

Figure S6: A schematic graph of the method for generating multiple pules. The interferometers are implemented by BSs. One arm of each interferometer is equipped with a delayer, introducing a $2^k$ units of time delay to the $k$th interferometer. We marked the first three and the last interferometers by (i), (ii), (iii), and (iv) for presenting the explicit description.

**References**

[1] N. Khaneja and S. J. Glaser, *Cartan Decomposition of SU(2n) and Control of Spin Systems*, Chemical Physics **267**, 11 (2001).

[2] V. V. Shende, S. S. Bullock, and I. L. Markov, *Synthesis of Quantum-Logic Circuits*, IEEE Trans. Comput.-Aided Des. Integr. Circuits Syst. **25**, 1000 (2006).

[3] M. Möttönen, J. J. Vartiainen, V. Bergholm, and M. M. Salomaa, *Quantum Circuits for General Multiqubit Gates*, Phys. Rev. Lett. **93**, 130502 (2004).

[4] C. C. Paige and M. Wei, *History and Generality of the CS Decomposition*, Linear Algebra and Its Applications **208–209**, 303 (1994).

[5] *Quantum Computational Networks*, Proc. R. Soc. Lond. A **425**, 73 (1989).

[6] D. P. DiVincenzo, *Two-Bit Gates Are Universal for Quantum Computation*, Phys. Rev. A **51**, 1015 (1995).

[7] A. Barenco, C. H. Bennett, R. Cleve, D. P. DiVincenzo, N. Margolus, P. Shor, T. Sleator, J. Smolin, and H. Weinfurter, *Elementary Gates for Quantum Computation*, Phys. Rev. A **52**, 3457 (1995).

[8] D. Aharonov, *A Simple Proof That Toffoli and Hadamard Are Quantum Universal*, 4 (n.d.).

[9] M. A. Nielsen and I. L. Chuang, *Quantum Computation and Quantum Information* (Cambridge University Press, Cambridge ; New York, 2000).